\documentclass[12pt]{article}
\usepackage{bezier}
\usepackage{amssymb}
\usepackage{epsfig}

\newcommand{\nc}{\newcommand}

\nc{{\betap}}{{\beta^{\,\prime}}}

\nc{\mlgraph}{{multiple-line graph }}
\nc{\mlgraphs}{{mul\-tiple\--line graphs }}
\nc{\Mlgraph}{{Multiple-line graph }}

\nc{\mldiagram}{{multiple-line diagram }}
\nc{\mldiagrams}{{multiple-line diagrams }}

\nc{\mlmoment}{{multiple-line moment}}
\nc{\mlmoments}{{multiple-line moments}}

\nc{\be}{\begin{equation}}
\nc{\ee}{\end{equation}}
\nc{\bea}{\begin{eqnarray}}
\nc{\eea}{\end{eqnarray}}
\nc{\bela}{\begin{eqnarray*}}
\nc{\eela}{\end{eqnarray*}}

\nc{\eqn}[1]{{(\ref{#1})}}
\nc{\cA}{{\cal A}}
\nc{\cB}{{\cal B}}
\nc{\cC}{{\cal C}}
\nc{\cD}{{\cal D}}
\nc{\cE}{{\cal E}}
\nc{\cF}{{\cal F}}
\nc{\cG}{{\cal G}}
\nc{\cH}{{\cal H}}
\nc{\cI}{{\cal I}}
\nc{\cJ}{{\cal J}}
\nc{\cK}{{\cal K}}
\nc{\cL}{{\cal L}}
\nc{\cM}{{\cal M}}
\nc{\cN}{{\cal N}}
\nc{\cO}{{\cal O}}
\nc{\cP}{{\cal P}}
\nc{\cQ}{{\cal Q}}
\nc{\cR}{{\cal R}}
\nc{\cS}{{\cal S}}
\nc{\cT}{{\cal T}}
\nc{\cU}{{\cal U}}
\nc{\cV}{{\cal V}}
\nc{\cW}{{\cal W}}
\nc{\cX}{{\cal X}}
\nc{\cY}{{\cal Y}}
\nc{\cZ}{{\cal Z}}

\nc{\simo}[1]{{\stackrel{#1}{\simeq}}}
\nc{\geqo}[1]{{\stackrel{#1}{\geq}}}
\nc{\geo}[1]{{\stackrel{#1}{>}}}
\nc{\guo}[1]{{\stackrel{#1}{\succ}}}

\nc{\rbo}{\raisebox}
\nc{\RR} {\rangle \! \rangle}
\nc{\LL} {\langle \! \langle}
\nc{\rmi}[1]{{\mbox{\small #1}}}
\nc{\eq}{eq.~}
\nc{\nr}[1]{(\ref{#1})}
\nc{\ul}{\underline}
\nc{\mc}{\multicolumn}
\nc{\todo}[1]{\par\noindent{\bf $\rightarrow$ #1}}

\nc{\cu}{{\cal u}}

\begin{document}


\begin{flushright} {\small $\begin{array}{ l } \mbox{SPhT-00/xx} \\
    \mbox{WUB--00--11} \\
    \mbox{HD--THEP--00--25} \end{array} $}
\end{flushright}

\begin{center}
{\LARGE
Dynamical Linked Cluster Expansions \\
for Spin Glasses\footnote{Invited contribution to
RECENT RESEARCH DEVELOPMENTS IN  STATSTICAL \\
PHYSICS,
ed.~by {\sl Transworld Research Network}
}
}
\end{center}

\begin{center}
{\Large \baselineskip 20pt
Hildegard~Meyer-Ortmanns$^{\, a\,}$\footnote{E-mail address:
ortmanns@theorie.physik.uni-wuppertal.de} $\,$
\\
$\qquad$ and $\qquad$
\\
Thomas Reisz$^{\, b,c\,}$\footnote{
E-mail address: reisz@thphys.uni-heidelberg.de,
supported by a Heisenberg Fellowship}
}
\end{center}

\centerline{$^a$ Institut f\"ur Theoretische Physik}
\centerline{Bergische Universit\"at Wuppertal}
\centerline{Gau\ss strasse 20}
\centerline{D-42097 Wuppertal, Germany}
\vspace{0.2cm}
\centerline{$^b$ Service de Physique Th{\'e}orique de Saclay}
\centerline{CE-Saclay}
\centerline{F-91191 Gif-sur Yvette Cedex, France}
\vspace{0.2cm}
\centerline{$^c$ Institut f\"ur Theoretische Physik}
\centerline{Universit\"at Heidelberg}
\centerline{Philosophenweg 16}
\centerline{D-69120 Heidelberg, Germany}


\begin{abstract}
Dynamical linked cluster expansions are linked cluster expansions with
hopping parameter terms endowed with their own dynamics. 
This amounts to a generalization from 2-point to point-link-point
interactions. An
associated graph theory with a generalized notion of connectivity
is reviewed.
We discuss physical applications to disordered systems,
in particular to spin glasses,
such as the bond-diluted Ising model and the Sherrington-Kirkpatrick
spin glass.
We derive the rules and identify the full set of graphs
that contribute to the series in the
quenched case.
This way it becomes possible to avoid the 
vague 
extrapolation from positive integer $n$ to $n=0$,
that usually goes along with an application of the replica trick.
\end{abstract}


%
%
%
%
\section{Introduction}

Linked cluster expansions (LCEs) have a long tradition in statistical
physics. Originally applied to classical fluids, later to magnetic systems
(\cite{wortis},\cite{itzykson},\cite{guttmann} and references therein),
they were generalized to applications in particle physics in the eighties
\cite{LW1}. There they have been used to study the
continuum limit of a lattice $\Phi^4$ field theory in 4 dimensions at zero 
temperature. In \cite{thomas1,thomas2}
they were further generalized to field
theories at finite temperature, simultaneously the highest 
order in the expansion parameter was increased to 18.
Usually the analytic expansions are obtained as graphical expansions.
Because of the
progress in computer facilities and the development of efficient
algorithms for generating the graphs, it is nowadays possible to handle
of the order of billions of graphs. The whole range
from high temperatures down to the critical region becomes available,
and thermodynamic quantities like critical indices and critical temperatures
are determined with high precision
(the precision is comparable or even better than in corresponding
high quality Monte Carlo results)
\cite{thomas2}-\cite{butera}.
An extension
of LCEs to a finite volume in combination with a high order in the 
expansion parameter turned out to be a particularly powerful tool for
investigating the phase structure of systems with first and second order
transitions by means of a finite size scaling analysis \cite{hilde1}.

Linked cluster expansions are series expansions of the free energy 
and connected correlation functions about an
ultralocal, decoupled theory in terms of a hopping parameter $K$.
The corresponding graphical representation is a sum in terms of connected
graphs. The value of $K$ parametrizes the strength of interactions
between fields at different lattice sites. Usually they are
chosen as nearest neighbours. In contrast to the ultralocal
terms of a generic interaction we will sometimes refer to hopping 
terms as non-ultralocal.

In this paper we develop dynamical linked cluster expansions (DLCEs).
These are linked cluster expansions with  hopping parameter terms
that are endowed with their own dynamics. 
Such systems are realized
in spin glasses with (fast) spins
and (slow) interactions \cite{sherrington}-\cite{penney}. 
They also occur in variational
estimates for SU(N)-gauge-Higgs systems, cf.~\cite{DLCE}.
Like LCEs they are expected to converge for a large class of interactions.

Formally DLCEs amount to a generalization of an expansion scheme 
from 2-point to point-link-point-interactions. 
These are interactions between
fields associated with two points and with one pair of
points called link. The points and links are not
necessarily embedded on a lattice, and the links need not be
restricted to nearest neighbours.
We have developed a new \mlgraph theory in which a generalized notion of
connectivity plays a central role.
Standard notions of equivalence classes of graphs like
1-line irreducible and 1-vertex irreducible graphs have been generalized,
and new notions like 1-multiple-line irreducible graphs were
defined in order to give a systematic classification.

The paper is organized as follows. In Sec.~2 we specify the
models that admit a DLCE. We introduce \mlgraphs 
and explain the idea behind the abstract notions of \mlgraph theory.
Detailed definitions of \mlgraphs, related notions and the
computation of weights are given in Sect.~3.
Sect.~4 treats the issue of renormalization in the sense of
suitable resummations of graphs.
Applications to spin glasses
are presented in Sect.~5. There it
is of particular interest that DLCEs allow for the possibility of 
avoiding the replica trick. 
In the quenched limit we derive
that DLCEs must be restricted to a subclass of the corresponding
graphical expansion, so-called quenched DLCEs (QDLCEs).
We also list some examples for models whose phase structure
is accessible to QDLCEs.
To these belongs in particular the bond diluted Ising model.
Sect.~6 contains the summary and conclusions.

%
%

%
%
\section{\label{models}A Short Primer to DLCEs}

In  this section we first specify the class of models for which we develop
dynamical linked cluster expansions. Next we illustrate some basic notions
of multiple-line graph theory, in particular the need for a new notion of
connectivity.

By $\Lambda_0$ we denote a finite or infinite set of points. One of its
realizations is a
hypercubic lattice in $D$ dimensions, 
infinite or finite in some directions with the topology of a torus.
$\Lambda_1$ denotes the set of unordered pairs $(x,y)$ of sites
$x,y\in\Lambda_0$, $x\not= y$, also called
unoriented links, and $\overline\Lambda_1$ a subset of $\Lambda_1$.

We consider physical systems with a partition function of the generic form
\bea\label{2.zgen}
  && Z(H,I,v) \; \equiv \; \exp{W(H,I,v)} \nonumber \\
  && \; = \;  \cN  \int \cD \phi \cD U
\exp{(-S(\phi,U,v))} \exp{(\sum_{x \in \Lambda_\circ}H(x)
          \phi(x) +\sum_{l \in \overline\Lambda_1}I(l)U(l))},
\eea
with measures
\be\label{2m}
    \cD\phi =  \prod_{x\in \Lambda_0}  d\phi(x)
\quad , \quad
 \cD U = \prod_{l \in \overline\Lambda_1} dU(l)
\ee
and action
\be\label{2act}
S(\phi,U,v) \; = \; \sum_{x \in \Lambda_\circ}S^\circ(\phi(x))
             +\sum_{l \in \overline\Lambda_1}S^1(U(l))
             - \frac{1}{2} \, \sum_{x,y \in \Lambda_\circ} 
               v(x,y)\phi(x)U(x,y)\phi(y) ,
\ee
with non-ultralocal couplings
\bea
 &&  v(x,y) \; = \; v(y,x) \; \not= \; 0 
 \qquad   \mbox{only for $(x,y)\in\overline\Lambda_1$}, \nonumber \\
 && \mbox{in particular} \; v(x,x) \; = \; 0.
\eea
For later convenience the normalization via $\cN$ is chosen such that
$W[0,0,0]=0$.

The field $\phi(x)$ is associated with the sites $x\in\Lambda_0$
and the field $U(l)$
lives on the links $l\in\overline\Lambda_1$,
and we write $U(x,y)=U(l)$ for $l=(x,y)$.
For definiteness and for simplicity of the notation here we assume
$\phi(x) \in {\bf R}$ and $U(l) \in {\bf R}$.
In our actual applications to spin glasses 
the $\phi$s are the (fast)
Ising spins and the $U$s $\in {\bf R}$ are the (slow) interactions.
The action is split into two ultralocal parts, $S^\circ$ depending
on fields on single sites, and $S^1$ depending on fields
on single links $l \in \overline\Lambda_1$.
For simplicity we choose $S^1$ as the same function for all
links $l\in\overline\Lambda_1$. 
We may identify $\overline\Lambda_1$ with the support of $v$,
\be
  \overline\Lambda_1 \; = \; \{ l=(x,y) \; \vert \;
  v(x,y) \not=0 \}.
\ee
The support of $v(x,y)$ need not be restricted to nearest neighbours,
also the precise form of $S^\circ$ and $S^1$
does not matter
for the generic description of DLCEs, $S^\circ$ and $S^1$ can be any
polynomials in $\phi$ and $U$, respectively.
The only restriction is 
the existence of the partition function.

Note that the
interaction term $v(x,y)~\phi(x)~U(x,y)~\phi(y)$ contains a
point-link-point-interaction and generalizes the 2-point-interactions 
$v(x,y)~\phi(x)~\phi(y)$ of usual hopping parameter expansions.
The effective coupling of the $\phi$ fields has its own dynamics
governed by $S^1(U)$, the reason why we have called our new
expansion scheme {\it dynamical} LCE.

Dynamical linked cluster expansions are induced from a Taylor expansion of
$W(H,I,v)=\ln Z(H,I,v)$ about $v=0$, the limit of a completely decoupled
system.
We want to express the series for $W$ in terms of connected graphs.
Let us consider the generating equation
\bea \label{2.genequation}
\partial W/\partial v(xy) & = & 1/2 <\phi(x) U(x,y) \phi(y)> \nonumber \\
      & = & 1/2 
   \biggl( W_{H(x) I(x,y) H(y)} + W_{H(x)H(y)}W_{I(x,y)} \nonumber \\
            & + & W_{H(x)I(x,y)}W_{H(y)}  \nonumber
         +W_{I(x,y)H(y)}W_{H(x)}  \nonumber   \\
      & + & W_{H(x)}W_{H(y)}W_{I(x,y)} 
   \biggr) .
\eea

Here $<\cdot>$ denotes the normalized expectation value
w.r.t. the partition function of Eq.~(\ref{2.zgen}). Subscripts $H(x)$ and
$I(x,y)=I(y,x)=I(l)$ denote the derivatives of $W$ w.r.t.
$H(x)$ and $I(x,y)$, respectively.

Next we would like to represent the right hand side
of Eq.~(\ref{2.genequation}) 
in terms of connected graphs.
Once we have such a representation for the first derivative of $W$ w.r.t. 
$v$, grapical expansions for the higher derivatives can be traced back
to the first one. 

For each $W$ in Eq.~(\ref{2.genequation}) we draw a shaded bubble,
for each derivative w.r.t. $H$ a solid line, called a $\phi$-line, 
with endpoint vertex $x$,
and for each derivative w.r.t. $I$ a dashed line,
called a $U$-line, with link label $l=(x,y)$. 
The main graphical constituents are
shown in Fig.~1. Two $\phi$-lines with endpoints $x$ and $y$ are
then joined by means of a dashed $U$-line with label $l$, 
if the link $l$ has $x$
and $y$ as its endpoints, i.e. $l=(x,y)$. According to these rules
Eq.~(\ref{2.genequation}),
multiplied by $v(x,y)$ and summed over $x$ and $y$,
is represented by Fig.~2.
Note that, because of the Taylor operation, each solid line from $x$
to $y$ carries a factor $v(x,y)$.

Since the actual need for a new type of connectivity is not 
quite obvious
from Fig.~2, because Eq.~(\ref{2.genequation}) does not contain higher
than first order derivatives w.r.t. $I$, let us consider a term

\begin{figure}[ht]

\begin{center}
\setlength{\unitlength}{0.8cm}

%
%
\begin{picture}(15.0,3.0)

%
%
%

\epsfig{bbllx=-333,bblly=324,
        bburx=947,bbury=684,
        file=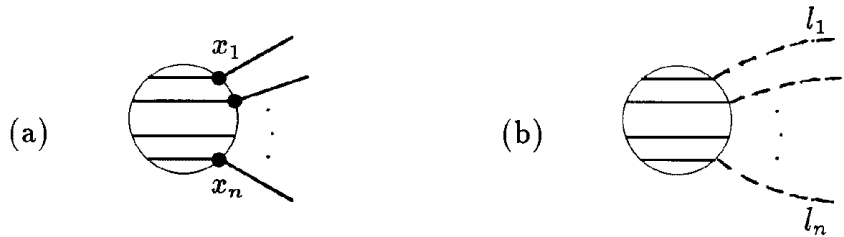,
        scale=0.27}

%
%

\end{picture}
%
%
\end{center}

\caption{\label{bubbles} Graphical representation of the derivatives
of $W(H,I,v)$.
(a) $n$-point function 
${\partial^n W}/{\partial H(x_1) \cdots \partial H(x_n)}$,
(b) $n$-link function
${\partial^n W}/{\partial I(l_1) \cdots \partial I(l_n)}$.
}
\end{figure}

\begin{figure}[ht]

\begin{center}
\setlength{\unitlength}{0.8cm}

%
%
\begin{picture}(15.0,7.0)

%
%
%

\epsfig{bbllx=-333,bblly=139,
        bburx=947,bbury=797,
        file=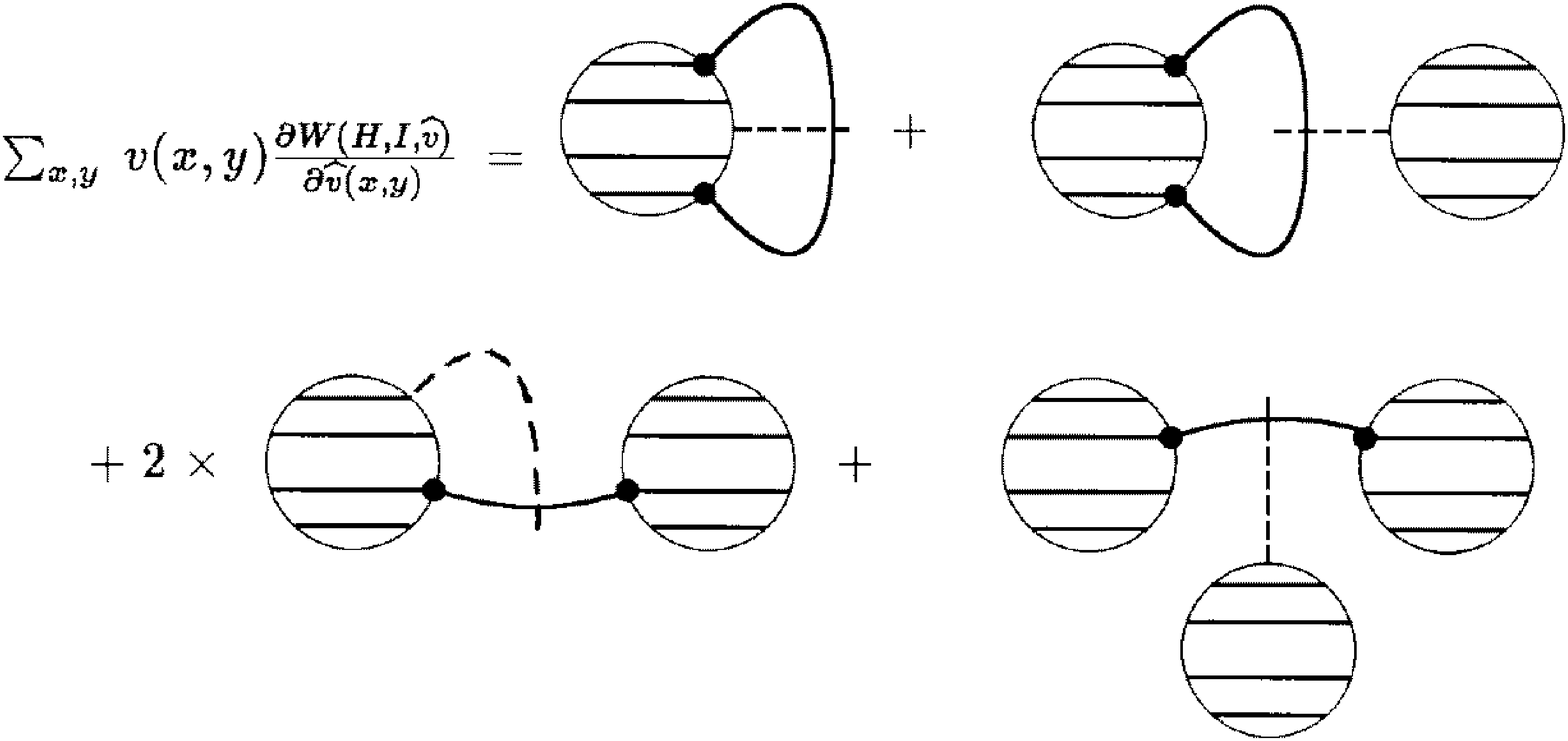,
        scale=0.27}

%
%

\end{picture}
%
%
\end{center}

\caption{\label{generation} Generating equation of the graphical
expansion of DLCEs.
The solid line in each graph carries a propagator $v(x,y)$.
A dashed $U$-line with label $l$ intersects a solid line with
endpoints $x$ and $y$ if $l=(x,y)$.}
\end{figure}

\be
W_{H(x)}W_{H(y)}W_{H(r)}W_{H(s)}
W_{I(x,y)I(r,s)}
\ee
occurring in the second derivative of $W$ w.r.t. $v(x,y)$, $v(r,s)$.
According to the above rules this term would be represented as shown in
Fig. 3a. While the 2 vertices in the last term of
Fig. 2 are connected
in the usual sense via a common (solid) line (the dashed line
with an attached bubble could be omitted in this case), the graph in 
Fig. 3a would be disconnected in the old sense,
since neither $x$ nor $y$ are line-connected with $r$ and $s$, but -as a
new feature of DLCE graphs- the lines from $x$ to $y$ and from $r$ to $s$
are connected via the dashed lines emerging from a common bubble shown
in the middle of the graph.
As we see from Fig. 3a, we need an additional notion of
connectivity referring to the possibility of multiple-line connectivity.
While the analytic expression is fixed, 
it is a matter of convenience
to further simplify the graphical notation of Fig.~3a 
at $v=0$. Two possibilities are shown in Fig.~3b
and Fig.~3c. To Fig. 3b we later refer in the formal definition
of the new type of multiple-line connectivity.
In the familiar standard notion of connectivity two vertices of a graph are
connected via lines. The vertices are line-connected.
Already there, in a dual language, one could call two
lines connected via vertices. The second formulation is just appropriate
for our need to define when two lines are connected. The corresponding
vertices mediating the connectivity of lines are visualized
by tubes, in Fig. 3b we have just one of them.
The tubes should be distinguished from the former type of
vertices represented as full dots which are connected via bare 
$\phi$-lines.
In Fig.~3c we show a simplified representation of Fig.~3b
that we actually use in graphical expansions.

\begin{figure}[h]

\begin{center}
\setlength{\unitlength}{0.8cm}

%
%
\begin{picture}(15.0,7.0)

%
%
%

\epsfig{bbllx=-333,bblly=186,
        bburx=947,bbury=606,
        file=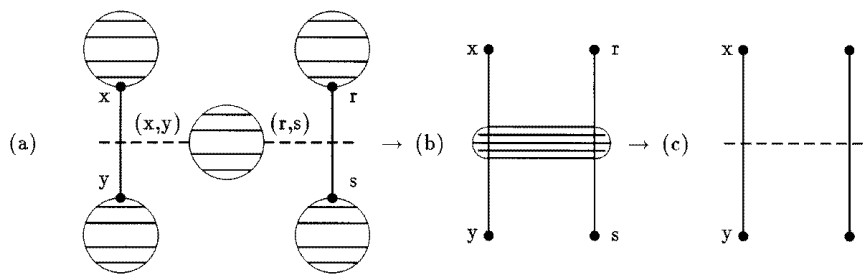,
        scale=0.27}

%
%

\end{picture}
%
%
\end{center}

\caption{\label{mlines} Representation of $W_{H(x)} W_{H(y)}
W_{H(r)} W_{H(s)} W_{I(x,y) I(r,s)}$.
(a) according to the rules of Fig.~1 and 2,
(b) same as (a), but at $v=0$ and simplified for a formal definition
of multiple-line connectivity, cf.~section 3,
(c) same as (b), but for use in the actual graphical
representations.}
\end{figure}

The derivative terms have to be evaluated at $v=0$. For $v=0$ we have
a decomposition of $W$ according to
\be
W(H,I,v=0)= \sum_{x \in \Lambda_0} W^\circ(H(x))
    +\sum_{l \in \overline\Lambda_1} W^1(I(l))
\ee
with 
\be\label{2.w0}
\exp{W^\circ(H)} \; \equiv \; Z^\circ(H) =
 \frac{\int^\infty_{-\infty} d\phi
 \exp{(-S^\circ(\phi)+H\phi)}}{\int^\infty_{-\infty} d\phi
 \exp{(-S^\circ(\phi))}}
\ee
and
\be\label{2.w1}
\exp{W^1(l)} \; \equiv \; Z^1(I) = \frac{\int^\infty_{-\infty} dU
\exp{(-S^1(U)+IU)}}{\int^\infty_{-\infty} dU
\exp{(-S^1(U))}} .
\ee
In Eq.s~(\ref{2.w0},\ref{2.w1}) we have omitted any single site or single
link dependence, because we assume that $S^\circ$ and 
$S^1$ are the same for all $x \in \Lambda_0$ and
$l\in \overline{\Lambda_1}$, respectively.
Therefore, at $v=0$, the only non-vanishing derivatives of $W$ are 
\be\label{2.wd1}
   \left. W_{H(x_1)H(x_2)...H(x_n)} \right\vert_{v=0} \; = \;
   \frac{\partial^n W^\circ(H(x_1))}
        {\partial H(x_1)^n} \cdot \delta_{x_1,x_2,...x_n}
\ee
and
\be\label{2.wd2}
   \left. W_{I(l_1)I(l_2)...I(l_m)} \right\vert_{v=0} \; = \;
   \frac{\partial^m W^1(I(l_1))}
        {\partial I(l_1)^m} \cdot \delta_{l_1,l_2,...l_m},
\ee
but mixed derivatives w.r.t.~H and I vanish.
As anticipated in Fig.s~3b and 3c, for $v=0$ we replace the dashed bubbles
and graphically distinguish between bubbles with $\phi$-lines
and bubbles with $U$-lines. We define
\be
  \left. 
{
\setlength{\unitlength}{0.8cm}
%
%
\begin{picture}(3.0,1.0)

%

\put(0.0,0.0){
\setlength{\unitlength}{0.8cm}
\begin{picture}(4.0,0.0)

\put(0.0,0.0){\circle*{0.16}}
\qbezier(0.0,0.0)(1.0,0.8)(2.0,0.8)
\qbezier(0.0,0.0)(1.0,0.5)(2.0,0.5)
\qbezier(0.0,0.0)(1.0,-0.7)(2.0,-0.7)

\put(1.2,0.2){\makebox(0.2,0){$\cdot$}}
\put(1.2,-0.1){\makebox(0.2,0){$\cdot$}}
\put(1.2,-0.4){\makebox(0.2,0){$\cdot$}}

\end{picture}
}
%
    
\end{picture}
}
    \right\} \; n
    \qquad = \; v^{\circ c}_n \; = \;
    \left( \frac{\partial^n W^\circ(H)}{\partial H^n} \right)_{H=0} 
\ee
for a connected n-point vertex with $n \ge 1$ bare $\phi$-lines
emerging from it and 
\be
  \left. 
{
\setlength{\unitlength}{0.8cm}
%
%
\begin{picture}(4.0,1.0)

%

\put(0.0,-0.6){
\setlength{\unitlength}{0.8cm}
\begin{picture}(4.0,0.0)

\qbezier(1.0,1.5)(2.0,2.1)(3.0,1.5)
\qbezier(1.0,1.0)(2.0,1.6)(3.0,1.0)
\qbezier(1.0,0.0)(2.0,0.6)(3.0,0.0)

\put(1.6,1.1){\makebox(0.2,0){$\cdot$}}
\put(1.6,0.8){\makebox(0.2,0){$\cdot$}}
\put(1.6,0.5){\makebox(0.2,0){$\cdot$}}

\put(2.0,1.85){\line(0,1){0.2}}
\put(2.0,1.6){\line(0,1){0.2}}
\put(2.0,1.35){\line(0,1){0.2}}
\put(2.0,1.1){\line(0,1){0.2}}
\put(2.0,0.85){\line(0,1){0.2}}
\put(2.0,0.6){\line(0,1){0.2}}
\put(2.0,0.35){\line(0,1){0.2}}
\put(2.0,0.1){\line(0,1){0.2}}

\end{picture}
}
%

\end{picture}
}
   \right\} \; \nu
   \qquad = \; m^{1 c}_\nu \; = \;
   \left( \frac{\partial^\nu W^1(I)}{\partial I^\nu} \right)_{I=0}
\ee
for a connected $\nu$-line consisting of $\nu$ bare lines.
If $\nu=1$, we often omit the dashed line. If the bare lines
of a $\nu$-line are internal $\phi$-lines,
they get vertices attached to their endpoints, if they are external $U$-lines,
no vertices will be attached.
 
\vskip10pt

Let $V$ denote the lattice volume in $D$ dimensions.
The Taylor expansion of $W$ about $v=0$ to 
second order in $v$ then reads
\bea\label{2.long}
W(H,I,v) & = & W(H,I,v=0) \nonumber \\
 &+&\sum_{x,y \in \Lambda_0} v(x,y) \, \frac{1}{2} \, W_{H(x)}
W_{H(y)}W_{I(x,y)}  \nonumber \\
 & + & \frac{1}{2} \, \sum_{x,y,r,s \in \Lambda_0} 
  \frac{1}{4} \; v(x,y) v(r,s) 
 \nonumber \\ 
 \cdot 
 & \biggl( & 
    4 \, W_{H(y)}W_{H(s)}W_{H(r)H(x)}W_{I(x,y)}W_{I(r,s)} \nonumber \\
  &+& 2 \, W_{H(x)H(r)}W_{H(y)H(s)}W_{I(x,y)}W_{I(r,s)}  \nonumber \\
  &+& 4 \, W_{H(y)}W_{H(s)}W_{H(r)H(x)}W_{I(x,y)I(r,s)}  \nonumber \\
  &+& 2 \, W_{H(r)H(x)}W_{H(y)H(s)}W_{I(r,y)I(x,s)} \nonumber \\
  &+& W_{H(x)}W_{H(y)}W_{H(r)}W_{H(s)}W_{I(x,y)I(r,s)}
  \biggr)_{v=0} \nonumber \\
 &+& O(v^3),
\eea
where we have used that $v(x,x)=0$. For each $W$ in the products of
$W$s we now insert Eq.s~(\ref{2.wd1}),(\ref{2.wd2}).

If we choose $v$ in a standard way as next-neighbour
couplings 
\be
  v(x,y) = 2K \sum_{\mu=0}^{D-1} (\delta_{x+\hat{\mu},y}+
\delta_{x-\hat{\mu},y})
\ee
with $\hat{\mu}$ denoting the unit vector in $\mu$-direction,
Eq.~(\ref{2.long}) becomes in a graphical representation at $H=I=0$
\bea\label{2.graph}
  \frac{W(0,0,v)}{V} & = & (2K) \;\;
   \frac{1}{2} \; (2D) \;
%
{
\setlength{\unitlength}{0.8cm}
\begin{picture}(4.0,1.0)

\put(0.0,0.0){
\setlength{\unitlength}{1.0cm}
\begin{picture}(4.0,1.0)

\put(0.0,0.2){\circle*{0.16}}
\put(2.0,0.2){\circle*{0.16}}
\put(0.0,0.2){\line(1,0){2.0}}

\end{picture}
}

\end{picture}
%
}
 \nonumber \\
 &+& (2K)^2 \;\; \biggl\{ \frac{1}{2} \; (2D)^2 \;
%
{
\setlength{\unitlength}{0.8cm}
\begin{picture}(3.5,1.0)

\put(0.0,0.0){
\setlength{\unitlength}{1.0cm}
\begin{picture}(4.0,1.0)

\put(0.0,-0.2){\circle*{0.16}}
\put(0.0,0.6){\circle*{0.16}}
\put(2.0,0.2){\circle*{0.16}}
\qbezier(0.0,-0.2)(1.0,-0.2)(2.0,0.2)
\qbezier(0.0,0.6)(1.0,0.6)(2.0,0.2)

\end{picture}
}

\end{picture}
%
}
 \; + \; \frac{1}{4} \; (2D)
%
{
\setlength{\unitlength}{0.8cm}
\begin{picture}(4.0,1.0)

\put(0.0,0.0){
\setlength{\unitlength}{1.0cm}
\begin{picture}(4.0,1.0)

\put(0.0,0.2){\circle*{0.16}}
\put(2.0,0.2){\circle*{0.16}}
\qbezier(0.0,0.2)(1.0,-0.4)(2.0,0.2)
\qbezier(0.0,0.2)(1.0,0.8)(2.0,0.2)

\end{picture}
}

\end{picture}
%
}
 \nonumber \\
 &\;& \qquad\quad + \frac{1}{2} \; (2D)^2 \;
%
{
\setlength{\unitlength}{0.8cm}
\begin{picture}(3.5,1.0)

\put(0.0,0.0){
\setlength{\unitlength}{1.0cm}
\begin{picture}(4.0,1.0)

\put(0.0,-0.2){\circle*{0.16}}
\put(0.0,0.6){\circle*{0.16}}
\put(2.0,0.2){\circle*{0.16}}
\qbezier(0.0,-0.2)(1.0,-0.2)(2.0,0.2)
\qbezier(0.0,0.6)(1.0,0.6)(2.0,0.2)

\put(0.9,0.2){\line(0,1){0.15}}
\put(0.9,0.4){\line(0,1){0.15}}
\put(0.9,0.6){\line(0,1){0.15}}
\put(0.9,0.0){\line(0,1){0.15}}
\put(0.9,-0.2){\line(0,1){0.15}}
\put(0.9,-0.4){\line(0,1){0.15}}

\end{picture}
}

\end{picture}
%
}
 \; + \; \frac{1}{4} \; (2D)
%
{
\setlength{\unitlength}{0.8cm}
\begin{picture}(4.0,1.0)

\put(0.0,0.0){
\setlength{\unitlength}{1.0cm}
\begin{picture}(4.0,1.0)

\put(0.0,0.2){\circle*{0.16}}
\put(2.0,0.2){\circle*{0.16}}
\qbezier(0.0,0.2)(1.0,-0.4)(2.0,0.2)
\qbezier(0.0,0.2)(1.0,0.8)(2.0,0.2)

\put(1.0,0.2){\line(0,1){0.15}}
\put(1.0,0.4){\line(0,1){0.15}}
\put(1.0,0.6){\line(0,1){0.15}}
\put(1.0,0.0){\line(0,1){0.15}}
\put(1.0,-0.2){\line(0,1){0.15}}
\put(1.0,-0.4){\line(0,1){0.15}}

\end{picture}
}

\end{picture}
%
}
 \\
 &\;& \qquad\quad + \frac{1}{8} \; 2(2D) \;
%
{
\setlength{\unitlength}{0.8cm}
\begin{picture}(3.5,1.0)

\put(0.0,0.0){
\setlength{\unitlength}{1.0cm}
\begin{picture}(4.0,1.0)

\put(0.0,0.0){\circle*{0.16}}
\put(0.0,0.3){\circle*{0.16}}
\put(2.0,0.0){\circle*{0.16}}
\put(2.0,0.3){\circle*{0.16}}
\qbezier(0.0,0.0)(1.0,-0.5)(2.0,0.0)
\qbezier(0.0,0.3)(1.0,0.8)(2.0,0.3)

\put(1.1,0.2){\line(0,1){0.15}}
\put(1.1,0.4){\line(0,1){0.15}}
\put(1.1,0.6){\line(0,1){0.15}}
\put(1.1,0.0){\line(0,1){0.15}}
\put(1.1,-0.2){\line(0,1){0.15}}
\put(1.1,-0.4){\line(0,1){0.15}}

\end{picture}
}

\end{picture}
%
}
 \biggr\}
 \nonumber \\
 & + & O(K^3) . \nonumber
\eea

For clarity, here we have written explicitly the topological 
symmetry factors and the lattice embedding numbers. 
(Usually graphs represent their full weights including these factors.)
Note that the first two graphs of the second order contribution
also occur in a usual LCE with frozen
$U$-dynamics, the next two differ by an additional dashed 2-line 
and the last one becomes even
disconnected without the dashed line.

As usual, graphical expansions for correlation functions, in particular
susceptibilities, are generated from $W(H,I,v)$ by taking derivatives
w.r.t. the external fields $H$ and $I$.
Graphically this amounts to attaching external
$\phi$-lines and $U$-lines with
\be
{
\setlength{\unitlength}{0.8cm}
%
%
\begin{picture}(16.0,3.0)


\put(0.0,2.0){
\setlength{\unitlength}{0.8cm}
\begin{picture}(4.0,0.0)

\qbezier(0.0,0.0)(1.0,1.0)(2.0,1.0)
\put(2.0,1.0){\circle*{0.16}}

\put(3.2,0.5){\makebox(8.0,0){(1 endpoint) attached to vertices, e.g.}}
%

\qbezier(13.0,0.5)(14.0,0.5)(15.0,0.5)
\qbezier(13.0,0.5)(14.0,1.2)(15.0,0.5)
\qbezier(13.0,0.5)(14.0,-0.2)(15.0,0.5)
\put(13.0,0.5){\circle*{0.16}}
\put(15.0,0.5){\circle*{0.16}}
\qbezier(15.0,0.5)(15.5,0.5)(16.0,0.5)
\qbezier(13.0,0.5)(12.5,0.5)(12.0,0.0)

\put(14.0,-0.1){\line(0,1){0.15}}
\put(14.0,0.1){\line(0,1){0.15}}
\put(14.0,0.3){\line(0,1){0.15}}
\put(14.0,0.5){\line(0,1){0.15}}

\end{picture}
}
%


\put(0.0,0.0){
\setlength{\unitlength}{0.8cm}
\begin{picture}(16.0,0.0)

\qbezier(0.0,0.0)(1.0,1.0)(2.0,0.0)

\put(3.2,0.5){\makebox(8.0,0){(no endpoint) attached to $\nu$-lines, e.g.}}
%

\qbezier(13.0,0.5)(14.0,0.5)(15.0,0.5)
\qbezier(13.0,0.5)(14.0,1.2)(15.0,0.5)
\qbezier(13.0,0.5)(14.0,-0.2)(15.0,0.5)
\put(13.0,0.5){\circle*{0.16}}
\put(15.0,0.5){\circle*{0.16}}

\qbezier(13.0,-0.3)(14.0,0.2)(15.0,-0.3)
\put(14.0,-0.3){\line(0,1){0.15}}
\put(14.0,-0.1){\line(0,1){0.15}}
\put(14.0,0.1){\line(0,1){0.15}}
\put(14.0,0.3){\line(0,1){0.15}}
\put(14.0,0.5){\line(0,1){0.15}}

\end{picture}
}
%

\end{picture}
}
\ee
In passing we remark that the conventional LCE is included as a special case
of the DLCE, if the $U$-dynamics is "frozen" to some value $U_0\neq 0$,
so that
\bea
   W^1(I) & = &-S_1(U_0)+ I U_0, \nonumber \\
   \frac{\partial W^1(I)}{\partial I} & = & U_0, \\
   \frac{\partial^n W^1(I)}{\partial I^n} & = & 0
    \quad\mbox{for all $n>1$},  \nonumber
\eea
i.e., no n-lines do occur with $n>1$. In this case it becomes redundant to
attach dashed lines to bare lines. 
As mentioned above, in an LCE only the first three
contributions would be left in Eq.~(\ref{2.graph}).
%
%
%
%
\section{\label{basics}Graphical expansion}

%
%

\subsection{\Mlgraph theory}

The definition of a \mlgraph as it will be given here is adapted
to the computation of susceptibilities
and the free energy,
where the sum is taken over all possible locations of the fields.
The definition easily generalizes to correlation functions.

For details of the standard definiton of graphs in the framework of
linked cluster expansions and related notions
we refer  e.g.~to \cite{LW1,thomas1}.
Here, for convenience, we briefly recall the very definition of a graph
to point out the new properties
of \mlgraphs as defined below in this section.

A (standard LCE) graph or diagram is a structure
\be
  \widetilde\Gamma = (\widetilde\cL_\Gamma,
   \widetilde\cB_\Gamma,\widetilde E_\Gamma,\widetilde\Phi_\Gamma),
\ee
where $\widetilde\cL_\Gamma$ and $\widetilde\cB_\Gamma\not=\emptyset$
are disjoint sets
of internal lines and vertices of $\widetilde\Gamma$, respectively.
$\widetilde E_\Gamma$ is a map
\bea
  \widetilde E_\Gamma: \widetilde\cB_\Gamma & \to & \{0,1,2,\ldots\},
  \nonumber \\
   v & \to & \widetilde E_\Gamma(v)
\eea
that assigns to every vertex $v$ the number of external lines 
$\widetilde E_\Gamma(v)$ attached to it.
Finally, $\widetilde\Phi_\Gamma$ is the incidence relation
that assigns internal lines
to their two endpoints.

A \mlgraph or \mldiagram is a structure
\be
  \Gamma = (\cL_\Gamma, \cM_\Gamma, \cB_\Gamma,
             E_\Gamma^{(\phi)}, E_\Gamma^{(U)},
             \Phi_\Gamma, \Psi_\Gamma).
\ee
$\cL_\Gamma$, $\cM_\Gamma$ and $\cB_\Gamma$
are three mutually disjoint sets,
\bea
    \cL_\Gamma  &=& \mbox{set of bare internal lines of $\Gamma$}, \\
    \cM_\Gamma  &=& \mbox{set of multiple lines of $\Gamma$}, \\
    \cB_\Gamma  &=& \mbox{set of vertices of $\Gamma$}.
\eea
$E_\Gamma^{(\phi)}$ is a map
\bea
  E_\Gamma^{(\phi)}: \cB_\Gamma & \to & \{0,1,2,\ldots\}, \nonumber \\
  v & \to & E_\Gamma^{(\phi)}(v)
\eea
that assigns to every vertex $v$ the number of bare external $\phi$-lines
$E_\Gamma^{(\phi)}(v)$ attached to $v$.
Every such $\phi$-line represents a field $\phi$.
The number of external $\phi$-lines of $\Gamma$ is denoted by
$E_\Gamma^{(\phi)}=\sum_{v\in\cB_\Gamma} E_\Gamma^{(\phi)}(v)$.
Similarly, $E_\Gamma^{(U)}$ is a map
\bea
  E_\Gamma^{(U)}: \cM_\Gamma & \to & \{0,1,2,\ldots\}, \nonumber \\
  m & \to & E_\Gamma^{(U)}(m)
\eea
that assigns to every multiple line $m$ the number of external $U$-lines
$E_\Gamma^{(U)}(m)$ attached to $m$.
Every such $U$-line represents a field $U$ associated with a lattice link.
The number of external $U$-lines of $\Gamma$ is given by
$E_\Gamma^{(U)}=\sum_{m\in\cM_\Gamma} E_\Gamma^{(U)}(m)$.

Furthermore, $\Phi_\Gamma$ and $\Psi_\Gamma$ are incidence 
relations that assign bare internal lines
to their endpoint vertices and to their
multiple lines, respectively.
We treat lines as unoriented.
The generalization to oriented lines is easily done.
More precisely, let
$\overline{(\cB_\Gamma\times\cB_\Gamma)}^{\,\prime}$
be the set of unordered pairs of vertices $(v,w)$
with $v,w\in\cB_\Gamma$, $v\not= w$.
(The bar implies unordered pairs,
the prime the exclusions of $(v,v)$, $v\in\cB_\Gamma$.)
As in the standard linked cluster expansion, self-lines are
excluded. Every bare internal line is then mapped onto its pair of
endpoints via
\be
  \Phi_\Gamma: \cL_\Gamma \to 
   \overline{(\cB_\Gamma\times\cB_\Gamma)}^{\,\prime}.
\ee
We say that $v$ and $w$ are the endpoint vertices of $l\in\cL_\Gamma$
if $\Phi_\Gamma(l)=(v,w)$. If there is such an
$l\in\cL_\Gamma$, $v$ and $w$ are called neighbours.
Similarly, $\Psi_\Gamma$ is a map
\bea
  \Psi_\Gamma: \cL_\Gamma & \to & \cM_\Gamma, \nonumber \\
     l & \to & \Psi_\Gamma(l)
\eea
that maps every bare internal line to a multiple line.
A multiple line $m\in\cM_\Gamma$ is composed of
bare internal lines $l\in\cL_\Gamma$ which 
belong to $m$ in the sense that
$\Psi_\Gamma(l)=m$.
$l_{\cM_\Gamma}(m)$ is the total number of bare internal lines
belonging to $m$.
With $\nu=l_{\cM_\Gamma}(m)+E_\Gamma^{(U)}(m)$,
$m$ is called a $\nu$-line.
We always require that $\nu\geq 1$.
On the other hand,
every bare internal line belongs to one and only one multiple line.
For simplicity we often identify a $1$-line with the only one
bare line that belongs to it.
\vskip7pt
Next we introduce some further notions that will be used later.
External vertices are vertices having external $\phi$-lines attached,
\be
  \cB_{\Gamma,ext} =  \{ v\in\cB_\Gamma \; \vert \;
   E_\Gamma^{(\phi)}(v)\not=0 \},
\ee
whereas internal vertices do not,
$\cB_{\Gamma,int}=\cB_\Gamma\setminus\cB_{\Gamma,ext}$.
Similarly, external multiple lines have external 
$U$-lines attached,
\be
  \cM_{\Gamma,ext} =  \{ m\in\cM_\Gamma \; \vert \;
   E_\Gamma^{(U)}(m)\not=0 \},
\ee
and the complement in $\cM_\Gamma$ are the internal multiple lines,
$\cM_{\Gamma,int}=\cM_\Gamma\setminus\cM_{\Gamma,ext}$.

For every pair of vertices $v,w\in\cB_\Gamma$, $v\not= w$,
let $\overline\Phi^1(v,w)$ be the set of lines with
endpoint vertices $v$ and $w$, and
$\vert\overline\Phi^1(v,w)\vert$
the number of these lines. Thus
$\overline\Phi^1(v,w)$ is the set of lines $v$ and $w$ have
in common.
With $E_\Gamma^{(\phi)}(v)$ denoting the number of
external $\phi$-lines attached to $v\in\cB_\Gamma$,
\be
   t_{\cB_\Gamma}(v) \; = \;
   \sum_{w\in\cB_\Gamma} \vert\overline\Phi^1(v,w)\vert
    \; + \; E_\Gamma^{(\phi)}(v)
\ee
is the total number of bare lines attached to $v$.

\vskip7pt

Some topological notions and global properties of graphs will be
of major interest in the following.
A central notion is the connectivity of a \mlgraph.
Recall that we want to consider the DLCE expansion of the free energy
and of truncated correlation functions as an expansion
in connected graphs.
As indicated in section 2,
the main generalization compared to the common notion of
connectivity of a graph which is required here is that an additional
type of connectivity is provided by multiple-lines.
To define the connectivity of a \mlgraph $\Gamma$,
$\Gamma$ first is mapped to a (standard) LCE graph $\overline{\Gamma}$
to which the standard notion of connectivity applies.
There are various equivalent ways to define such a map.
We choose the following one.

\begin{itemize}

\item
For every multiple-line $m\in\cM_\Gamma$
define a new vertex $w(m)$. Let
$\widetilde\cB_\Gamma = \{ w(m) \vert m\in\cM_\Gamma \}$
and define
$\overline\cB = \cB_\Gamma \cup \widetilde\cB_\Gamma$
as the union of the vertices of $\Gamma$ and the new set of
vertices originating from the multiple-lines.

\item
For every bare internal line $l\in\cL_\Gamma$
define two new internal lines
$l_1,l_2$ and incidence relations
\bea
   \overline\Phi(l_1) & = & (v_1, w(\Psi_\Gamma(l))), \nonumber \\
   \overline\Phi(l_2) & = & (v_2, w(\Psi_\Gamma(l))),
\eea
where $v_1$ and $v_2$ are the two endpoint vertices of $l$.
The set of all lines
$l_1,l_2$, for all $l\in\cL_\Gamma$,
is denoted by $\overline\cL$.

\item
Define the external incidence relations
\bea
  \overline{E}\quad:\quad \overline\cB & \to & \{0,1,2,\ldots\}, \nonumber \\
      \overline{E}(v) & = & E_\Gamma^{(\phi)}(v),
        \quad \mbox{for $v\in\cB_\Gamma$}, \\
     \overline{E}(v) & = & E_\Gamma^{(U)}(m),
        \quad \mbox{for $v=w(m)\in\widetilde\cB_\Gamma$}.
    \nonumber
\eea

\end{itemize}

Now, $\overline\Gamma$ is defined by
\be
    \overline\Gamma \; = \; (\overline\cL,
    \overline\cB, \overline{E}, \overline\Phi ).
\ee

Having defined the standard LCE graph $\overline\Gamma$
for any \mlgraph $\Gamma$, we call
$\Gamma$ multiple-line connected or just connected 
if $\overline\Gamma$ is connected
(in the usual sense).
In Fig.~4 we have given two examples for a connected (upper graph)
and a disconnected (lower graph) \mlgraph.

\begin{figure}[h]

\begin{center}
\setlength{\unitlength}{0.8cm}

%
%
\begin{picture}(15.0,7.0)

%
%

\epsfig{bbllx=-333,bblly=50,
        bburx=947,bbury=888,
        file=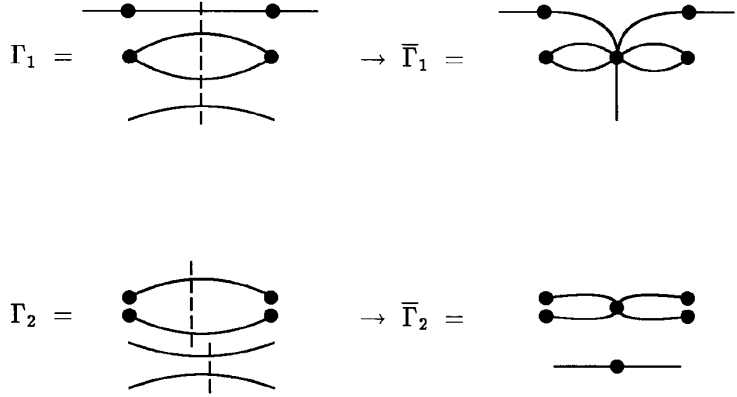,
        scale=0.22}

%
%

\end{picture}
%
%
\end{center}

\caption{\label{connected} Example of multiple-line connectivity.
The upper \mlgraph $\Gamma_1$ is connected because the graph
$\overline\Gamma_1$ is connected in the conventional sense.
The lower \mlgraph $\Gamma_2$ is disconnected because
$\overline\Gamma_2$ is so.  
}
\end{figure}

\vskip7pt

The next important notion is the topological equivalence of two
\mlgraphs.
Two \mlgraphs
\be \label{top.equiv}
  \Gamma_i \; = \; (\cL_i, \cM_i, \cB_i,
   E^{(\phi)}_i, E^{(U)}_i, \Phi_i, \Psi_i), \quad
   i=1,2
\ee
are called (topologically) equivalent if there are three
invertible maps
\bea
   \phi_1:\cB_1 &\to& \cB_2 , \nonumber \\
   \phi_2:\cL_1 &\to& \cL_2, \\
   \phi_3:\cM_1 &\to& \cM_2, \nonumber
\eea
such that
\bea
  \Phi_2 \circ \phi_2 &=& \overline\phi_1 \circ \Phi_1, \nonumber \\
  \Psi_2 \circ \phi_2 &=& \phi_3 \circ \Psi_1,
\eea
and
\bea
  E_2^{(\phi)} \circ \phi_1 &=& E_1^{(\phi)}, \nonumber \\
  E_2^{(U)} \circ \phi_3 &=& E_1^{(U)}.
\eea
Here $\circ$ means decomposition of maps, and
\bea 
  \overline\phi_1: \overline{\cB_1\times\cB_1}^{\,\prime} & \to &
      \overline{\cB_2\times\cB_2}^{\,\prime} \nonumber \\
   \overline\phi_1(v,w)  & = & (\phi_1(v),\phi_1(w)).
\eea

A symmetry of a \mlgraph
$\Gamma = (\cL,\cM,\cB,E^{(\phi)},E^{(U)},\Phi,\Psi)$
is a triple of maps
$\phi_1:\cB\to\cB$, $\phi_2:\cL\to \cL$
and $\phi_3:\cM\to\cM$
such that
\bea 
  \Phi \circ \phi_2 &=& \overline\phi_1 \circ \Phi, \nonumber \\
  \Psi \circ \phi_2 &=& \phi_3 \circ \Psi,
\eea
and
\bea
  E^{(\phi)} \circ \phi_1 &=& E^{(\phi)} \nonumber \\
  E^{(U)} \circ \phi_3 &=& E^{(U)}.
\eea
The number of these maps is called the symmetry number of $\Gamma$.

We denote by $\cG_{E_1,E_2}(L)$ the set of equivalence classes
of connected \mlgraphs with $L$ bare internal lines,
$E_1$ external $\phi$-lines and $E_2$ external $U$-lines.
Furthermore we set
\be
  \cG_{E_1,E_2} \; := \; \bigcup\limits_{L\geq 0} \; \cG_{E_1,E_2}(L).
\ee

A multiple line graph $\Gamma$ does not need to have a vertex.
If $\cB_\Gamma =0$, we have $\cL_\Gamma =0$ as well.
If in addition $\Gamma$ is connected, $\cM_\Gamma$ consists of only one element,
with all external $U$-lines attached to it.
(We anticipate that $\Gamma$ is 1-multiple-line irreducible (1MLI) by definition.
For the definition of 1MLI cf.~section 4 below.)
The only graph of $\cG_{0,E}(L=0)$ is given by
\be
  \Gamma \; = \; \left.
{
\setlength{\unitlength}{0.8cm}
%
%
\begin{picture}(4.0,1.0)

%
%
\put(0.0,0.0){
\setlength{\unitlength}{1.0cm}
\begin{picture}(10.0,0.0)

\qbezier(0.9,0.6)(1.5,0.5)(2.1,0.6)
\qbezier(0.9,0.4)(1.5,0.3)(2.1,0.4)
\qbezier(0.9,-0.5)(1.5,-0.4)(2.1,-0.5)
\qbezier(0.9,-0.3)(1.5,-0.2)(2.1,-0.3)
\put(1.5,-0.55){\line(0,1){0.15}}
\put(1.5,-0.35){\line(0,1){0.15}}
\put(1.5,-0.15){\line(0,1){0.15}}
\put(1.5,0.05){\line(0,1){0.15}}
\put(1.5,0.25){\line(0,1){0.15}}
\put(1.5,0.45){\line(0,1){0.15}}
\put(1.5,0.65){\line(0,1){0.15}}
\put(1.2,0.15){\makebox(0.2,0){$\cdot$}}
\put(1.2,-0.05){\makebox(0.2,0){$\cdot$}}
\put(1.2,0.05){\makebox(0.2,0){$\cdot$}}

\end{picture}
}
%
%

\end{picture}
%
%
}
   \right\} \; E .
\ee
It represents the leading term of the susceptibility
\be \label{susc.0E}
  \chi_{0,E} \; = \;
  \frac{1}{V D} \sum_{l_1,\dots ,l_E \in\overline\Lambda_1}
    < U(l_1) \cdots U(l_E) >^c
\ee
and is given by
$\left.{\partial^E W^1(I)}/{\partial I^E}\right\vert_{I=0}$. 
The index $c$ in (\ref{susc.0E})
stands for truncated (connected) correlation.

\vskip7pt

By removal of a $\nu$-line $m\in\cM_\Gamma$ we mean that $m$ is
dropped together with all bare internal lines and all external
$U$-lines that belong to $m$.
This notion is explained in Fig.~5a.
(It is used in section 4 for $1$-lines to define
1-particle irreducible (1PI) and 
1-line irreducible (1LI) \mlgraphs.)

On the other hand, by decomposition of a $\nu$-line $m\in\cM_\Gamma$
we mean that $m$ is dropped together with the external $U$-lines
of $m$, but all bare internal lines that belong to $m$
are kept in the graph, being identified now with $1$-lines.
This notion will be used below to define
1MLI and renormalized \mlmoments .
It is illustrated in Fig.~5b.

Similarly, decomposition of a vertex $v\in\cB_\Gamma$
means to remove the vertex $v$ and to attach the free end of every line
that entered $v$ before to a new vertex, a separate one for each line.
This notion is used to define 1-vertex-irreducible (1VI)
and renormalized vertex moments
for \mlgraphs. For an example see Fig.~5c.

\begin{figure}[h]

\begin{center}
\setlength{\unitlength}{0.8cm}

%
%
\begin{picture}(15.0,7.0)

%
%

\epsfig{bbllx=-333,bblly=60,
        bburx=947,bbury=878,
        file=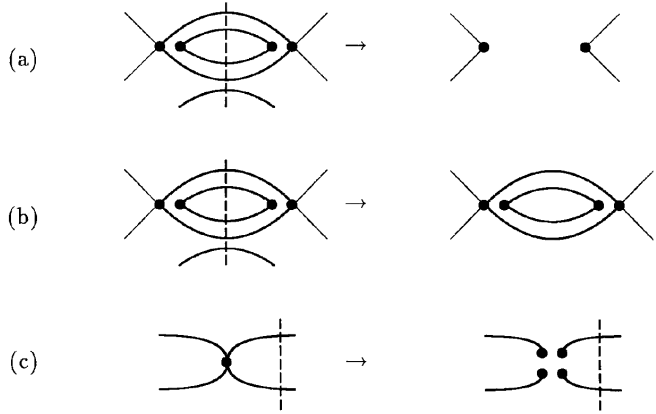,
        scale=0.20}

%
%

\end{picture}
%
%
\end{center}

\caption{\label{removdecomp} Removal (a) and decomposition (b) of
a $5$-line. Decomposition (c) of a vertex.}
\end{figure}

%
%

\subsection{Susceptibilities and weights}

In the last section
we have defined \mlgraphs and the notions of connectivity and 
equivalence of such graphs.
The definition is chosen in such a way 
that the series expansions
of the free energy and of truncated correlation functions are
obtained as a sum over equivalence classes of 
connected \mlgraphs.
The number $L$ of bare internal lines of a \mlgraph $\Gamma$
counts the order in the expansion parameter $v(x,y)$ to which
$\Gamma$ contributes.
If $v(x,y)$ is of the form
\be \label{susc.vform}
   v(x,y) \; = \; 2K \sum_{z\in\cN(x)} \delta_{y,z},
\ee
with $\cN(x)$ any finite
$x$-dependent set of lattice sites,
the contribution of $\Gamma$ is a multiple of $(2K)^L$.
Often used special cases are the nearest neighbour interactions
\be \label{susc.nn}
   v(x,y) \; = \; 2K \sum_{\mu=0}^{D-1}
   \left( \delta_{x,y+\widehat\mu} + \delta_{x,y-\widehat\mu}
   \right)
\ee
and the uniform interaction
\be \label{susc.mf}
   v(x,y) \; = \; 2K \; \left( 1 - \delta_{x,y} \right),
\ee
which is used in models of spin glasses and
partially annealed neural networks.
\vskip7pt
Susceptibilities of the $\phi$
and $U$ fields will be represented as
\bea \label{susc.susc}
  \chi_{E_1,E_2} & = &
  \frac{1}{V D} \sum_{x_1,\dots ,x_{E_1} \in \Lambda_0}
  \sum_{l_1,\dots ,l_{E_2} \in \overline\Lambda_1}
  < \phi(x_1) \cdots \phi(x_{E_1}) \;
    U(l_1) \cdots U(l_{E_2}) >^c \nonumber \\
  & \equiv & \frac{1}{V D} \sum_{x_1,\dots ,x_{E_1} \in \Lambda_0}
  \sum_{l_1,\dots ,l_{E_2} \in \overline\Lambda_1}
  \left .\frac{\partial^{E_1+E_2}W(H,I,v)}
       {\partial H(x_1)\cdots\partial H(x_{E_1})
        \partial I(l_1)\cdots\partial I(l_{E_2})}
  \right\vert_{H=I=0} \nonumber \\
   & = & \sum_{L\geq 0} (2K)^L 
  \sum_{\Gamma\in\cG_{E_1,E_2}(L)} w(\Gamma)
\eea
with lattice volume $V$ and dimension $D$.
Similar representations hold for higher moments $\mu$.

The weight $w(\Gamma)$ of a \mlgraph 
$\Gamma\in\cG_{E_1,E_2}(L)$ is given as the product of the
following factors

\begin{itemize}

\item
for every vertex $v\in\cB_\Gamma$ a factor
\be
   v^{\circ c}_n \; = \;
  \biggl( \frac{\partial^n W(H)^\circ}{\partial H^n} \biggr)_{H=0} ,
\ee
where $n=t_{\cB_\Gamma}(v)$ is the total number of bare lines
attached to $v$.

\item
for every multiple line $m\in\cM_\Gamma$ a factor
\be \label{susc.m1c}
   m^{1 c}_\nu \; = \;
 \biggl(  \frac{\partial^\nu W^1(I)}{\partial I^\nu} \biggr)_{I=0},
\ee
where $\nu=l_{\cM_\Gamma}(m)+E_\Gamma^{(U)}(m)$, that is
$m$ is a $\nu$-line,

\item
a factor $1/S_\Gamma$, where $S_\Gamma$ is the topological
symmetry number of $\Gamma$,

\item
a factor counting the permutation symmetry of external $\phi$-lines,
\be
   \frac{ E_\Gamma^{(\phi)}! }
   { \prod_{v\in\cB_\Gamma} E_\Gamma^{(\phi)}(v)! }  \quad ,
\ee

\item
a factor counting the permutation symmetry of external $U$-lines,
\be
   \frac{ E_\Gamma^{(U)}! }
   { \prod_{m\in\cM_\Gamma} E_\Gamma^{(U)}(m)! }  \quad ,
\ee

\item
the lattice embedding number of $\Gamma$, which is
the number of ways $\Gamma$ can be embedded on a lattice of
given geometry, e.~g.~on a hypercubic lattice.
To this end, the vertices of $\Gamma$ (if any) are placed onto lattice sites.
One arbitrary vertex is placed at a fixed lattice site, in order to
account for the volume factor $1/V$ in (\ref{susc.susc}).
A priori there is no exclusion principle.
This means that any number of vertices can be placed at the
same lattice site.
(This is sometimes called free embedding.)
Two restrictions apply to the embeddings.
The first constraint results from the fact that
a bare internal line represents a hopping propagator $v(x,y)$, with
lattice sites $x$ and $y$ at which the two endpoint vertices
of the line are placed at.
A reasonable computation of the embedding number takes into
account the particular form of $v(x,y)$ from the very beginning.
The second constraint is that
bare lines of the same multiple-line have to be mapped
on the same pair of sites.

For example, if $v(x,y)$ is the nearest neighbour interaction
(\ref{susc.nn}), two vertices which have 
at least one line in common are to be
placed at nearest neighbour lattice sites.
On the other hand, a propagator $v(x,y)$ of the form
(\ref{susc.mf}) implies a rather weak constraint in that
$x$ and $y$ must be different, but otherwise can be freely placed
over the lattice.

\end{itemize}

We remark that in case of a non-trivial internal symmetry
(such as considered in section 7) the expressions of
Eq.s~(\ref{susc.susc})-(\ref{susc.m1c})
must be modified appropriately. In particular,
the weight (\ref{susc.m1c}) of a multiple-line 
does no longer take such a simple form.
%
%
%
%

\section{Renormalization}

Truncated correlation functions, susceptibilities
and other moments are obtained
as sums over \mlgraphs that are connected.
Their number rapidly grows with increasing order, that is with
increasing number of bare internal lines.
The procedure of "renormalization" means that the connected moments 
are represented
in terms of reduced ones. The reduced moments are obtained by summation
over \mlgraph classes which are more restricted than just by their
property of being connected.
Of course the number of graphs of such classes is smaller.
Only the most restricted \mlgraph classes must be constructed.
The subsequent steps towards the moment computation are most
conveniently done by operating analytically with the reduced
moments.
In particular, it is no longer necessary to generate all connected
and the corresponding intermediate \mlgraph classes.

A connected \mlgraph $\Gamma$ is called 1-particle irreducible (1PI) if
it satisfies the following condition.
Remove an arbitrary $1$-line of $\Gamma$. There is at most one
connected component left that has external lines attached.
(This notion is the same as the one used in the context of 
Feynman graphs.)
On the other hand, if in addition the remaining graph is still
connected, then $\Gamma$ is called 1-line irreducible (1LI).
In many cases it is sufficient to use only the second notion.
It is for instance sufficient that all vertices are constrained to have
only an even number of lines attached, or more generally, if
graphs and subgraphs with one external line are forbidden.
For notational simplicity we assume in the
following that this is the case and henceforth refer only to the
notion 1LI
\footnote{In ref.~\cite{LW1,thomas1} the term 1PI was used instead.}.
The generalization to the case in which 1LI and 1PI graphs must be
distinguished goes along the same lines
as for LCEs, which was discussed in \cite{thomas3}.

By $\cG_{E_1,E_2}^{1LI}(L)$ we denote the subset of \mlgraphs
$\Gamma\in\cG_{E_1,E_2}(L)$ that are 1LI.
1LI-susceptibilities are defined as series in the hopping
parameter similarly as in (\ref{susc.susc})
by restricting the summation to 1LI graphs,
\be
  \chi_{E_1,E_2}^{1LI} =
   \sum_{L\geq 0} (2K)^L 
  \sum_{\Gamma\in\cG_{E_1,E_2}^{1LI}(L)} w(\Gamma) .
\ee
Susceptibilities are easily obtained in a closed form in terms of
1LI-susceptibilities $\chi^{1LI}$. It can be shown that the $\chi^{1LI}$s
can be obtained by an appropriate Legendre transform.
For instance
\bea
  \chi_{2,0} & = & \frac{\chi_{2,0}^{1LI}}
    {1 - \widetilde{v}(0) \chi_{2,0}^{1LI}} , \nonumber \\
   \chi_{2,1} & = & \frac{\chi_{2,1}^{1LI}}
    {(1 - \widetilde{v}(0) \chi_{2,0}^{1LI})^2} ,
\eea
where $\widetilde{v}(k)$ is the Fourier transform of the hopping
propagator $v(x,y)$,
\be
  v(x,y) \; = \; \int_{-\pi}^\pi \frac{d^Dk}{(2\pi)^D} \;
   e^{-ik\cdot (x-y)} \; \widetilde{v}(k) .
\ee

In LCEs the second important resummation comes from so called
vertex renormalizations.
This means partial resummation of graphs with specific properties
such as having one external vertex only. These sums then are
considered as "renormalized vertices" replacing the vertices
of graphs with complementary properties.
The procedure naturally leads to the notion of 1-vertex irreducibility
(1VI) and renormalized moments.

In DLCE we follow this procedure. The very definition of 1VI has
to be modified slightly for \mlgraphs because of the enhanced connectivity
properties due to multiple-lines.
In addition, as a natural generalization,
we supplement vertex renormalization by multiple-line renormalization.

A \mlgraph $\Gamma$ is called 1-vertex irreducible (1VI) if it
satisfies the following condition.
Decompose an arbitrary vertex $v\in\cB_\Gamma$.
Every connected component of the remaining graph has then at least one
external line attached. It can be a $\phi$-line or a $U$-line.
We write
\be
   \cG_{E_1,E_2}^{1VI}(L) \; = \;
   \{ \Gamma\in\cG_{E_1,E_2}^{1LI}(L) \; \vert \;
     \mbox{$\Gamma$ is 1VI} \}
\ee
for the set of equivalence classes of graphs that are both 1LI and 1VI,
with $E_1$ external $\phi$-lines, $E_2$ external $U$-lines
and $L$ bare internal lines.

The renormalized vertex moment graphs are 1LI graphs that have
precisely one external vertex and no external multiple line,
\be
  Q_k(L) \; = \;
  \{ \Gamma\in\cG_{k,0}^{1LI}(L) \; \vert \;
     \mbox{there is $v\in\cB_\Gamma$ with $E_\Gamma^{(\phi)}(v)=k$} \}.
\ee

A \mlgraph $\Gamma$ is called 1-multiple-line irreducible (1MLI) if
it satisfies the following criterion.
Decompose an arbitrary multiple-line $m\in\cM_\Gamma$.
Every remaining connected component has then at least one external line
attached. It can be a $\phi$-line or a $U$-line.
We write
\be
   \cG_{E_1,E_2}^{1MLI}(L) \; = \;
   \{ \Gamma\in\cG_{E_1,E_2}^{1LI}(L) \; \vert \;
     \mbox{$\Gamma$ is 1MLI} \}.
\ee
The renormalized multiple-line moment graphs are graphs that are
1LI and have precisely one external multiple-line, but no external
vertex,
\be
  R_k(L) \; = \;
  \{ \Gamma\in\cG_{0,k}^{1LI}(L) \; \vert \;
     \mbox{there is $m\in\cM_\Gamma$ with $E_\Gamma^{(U)}(m)=k$} \}.
\ee

The equivalence classes of graphs that are both 1VI and 1MLI
are denoted by
\be
   S_{E_1,E_2}(L) \; = \;
      \cG_{E_1,E_2}^{1VI}(L) \cap \cG_{E_1,E_2}^{1MLI}(L).
\ee

With the renormalized moment graphs as defined above, the
1LI-susceptibilities are now obtained in the form
\be
  \chi_{E_1,E_2}^{1LI} =
   \sum_{L\geq 0} (2K)^L 
  \sum_{\Gamma\in\cS_{E_1,E_2}(L)} \widetilde{w}(\Gamma) .
\ee
The weights $\widetilde{w}(\Gamma)$ are given as a product
of factors as described in the last subsection, with
the following two exceptions.

\begin{itemize}

\item
The vertex coupling constants $v^{\circ c}_n$ are replaced
by the renormalized vertex moments
\be
  v^{\circ c}_n \; \to \;
  v^c_n = \sum_{L\geq 0} (2K)^L
  \sum_{\Gamma\in Q_n(L)} w(\Gamma) .
\ee

\item
The multiple line coupling constants $m^{1 c}_\nu$ are replaced
by the renormalized multiple line moments
\be
  m^{1 c}_\nu \; \to \;
  m^c_\nu = \sum_{L\geq 0} (2K)^L
  \sum_{\Gamma\in R_\nu (L)} w(\Gamma) .
\ee

\end{itemize}

In the series representations above, the $w(\Gamma)$
are computed according to the rules
of subsection 3.2.

In \cite{DLCE} we have described an algorithmic construction
of graphs that is the first step for an automatic computer
aided generation.
In exceptional cases DLCEs can be summed up in a closed form.
Otherwise, a computer aided generation of graphs is unavoidable,
because the proliferation of DLCE graphs is pronounced even compared
to LCE graphs.
For LCE graphs we know that billions of graphs must be
included to obtain the critical exponents with the accuracy of 
some per mil.

%
%
%
%
\section{Applications to spin glasses}

In this section we consider applications of DLCEs to disordered
systems, in particular to spin glasses with "`slow"' interactions
coupled to "`fast"' spins. 
The interactions $J$ are assumed to be in equilibrium with a thermal
heat bath of inverse temperature $\betap$, 
while the spins $\sigma$ are equilibrated according to a second
inverse temperature $\beta$.
Both systems need not be mutually in equilibrium.
Let $Z_\beta(J)$ be the partition function that describes the
equilibrium distribution of the spins for given $J$s,
\be\label{neur.zspin}
  Z_\beta(J) \; = \; \sum_{\{ \sigma_i = \pm 1\}}
  \exp{(\beta\sum_{i<j} J_{(i,j)} \sigma_i \sigma_j }) .
\ee
The sum runs over pairs $(i,j)$ that need not be restricted
to nearest neighbours only. 

We further assume that the dynamics of the time evolution of the
slow interactions $J$ is governed by a Langevin equation
\be
   N \frac{d}{dt} J_{(i,j)} \ = \; 
   - \ \frac{\partial}{\partial J_{(i,j)}} \cH(J) \ + \ 
    \sqrt{N} \eta_{ij}(t)
\ee
with
\be\label{neur.h}
 \cH(J) \; = \; - \frac{1}{\beta} \ln{Z_\beta(J)}
 + \frac{1}{2}\mu N \sum_{i<j=1}^N J_{(i,j)}^2 .
\ee
Here $Z_\beta(J)$ is given by (\ref{neur.zspin}),
$N$ is the total number of spins, $\mu$ is a positive constant
and $\eta_{ij}$ is a stochastic gaussian white noise of zero mean
and correlation
\be
   < \eta_{ij}(t) \eta_{kl}(t^{\,\prime}) > \; = \;
    \frac{2}{\betap} \delta_{(ij),(kl)}
     \delta(t-t^{\,\prime}) .
\ee
Such a Langevin equation for the $J$s can be derived from an ansatz
which is motivated by neural networks
\cite{sherrington}-\cite{penney}.
Moreover, since the time evolution of the $J$s is determined by
a dissipative Langevin equation, the equilibrium distribution
of the slow variables is again a Boltzmann distribution, 
governed now by the second temperature ${\betap}^{-1}$,
\be\label{neur.z}
  Z_{\beta^\prime}^{\prime} \; = \; \cN \,
  \int_{-\infty}^\infty \prod_{i<j = 1}^N dJ_{(i,j)} \;
  \exp{(-\beta^\prime \cH(J))},
\ee
with $\cN$ some normalization that will be specified below.
The effective Hamiltonian $\cH$ of
$J$ is given by (\ref{neur.h}).

It is these equilibrium aspects of coupled systems of fast spins
and slow interactions that we can treat analytically with
DLCEs, as we will show below.

\vskip 5pt

Let us first rewrite $Z^\prime_{\beta^\prime}$ in the form
\bea\label{neur.zre}
  Z_{x\beta}^{\prime} & = &
  \int_{-\infty}^\infty \prod_{i<j = 1}^N 
  \left( \sqrt{\frac{QN}{2\pi}} dJ_{(i,j)} \right) \cdot
  \exp{(-\frac{1}{2} Q N \sum_{i<j} J_{(i,j)}^2})
  \;\; Z_\beta(J)^x \nonumber \\
  && \nonumber \\
  & \equiv & [[ Z_\beta(J)^x ]] ,
\eea
where we have introduced $Q=\beta\mu$ and real
$x=\beta^\prime/\beta$ as the ratio of two temperatures.
The normalization has been chosen such that 
$[[1]] = 1$.
In the limit of $x\to 0$ for fixed $\beta$, i.e.~$\betap\to 0$,
we have a quenched system.
The $J$ only feel the infinitely high temperature
${\betap}^{-1}$, but are decoupled from the spins.
The time scale of fluctuations of the spins is assumed to be so
short that the $J$ are only sensitive to averages of the
$\sigma$. Vice versa, the spin dynamics does depend on the $J$s.
Therefore the quantity of physical interest is not 
\be
   \ln 
   \biggl\lbrack
    \int_{-\infty}^\infty \prod_{i<j = 1}^N 
      \left( \sqrt{\frac{QN}{2\pi}} dJ_{(i,j)} \right)
      \cdot 
    \biggl( \sum_{\{\sigma_i\}} 
      \exp{( \beta \sum_{i<j=1}^N J_{(i,j)}
                      \sigma_i \sigma_j
          )}
    \biggr)^x
  \cdot  \exp{(-\frac{1}{2} Q N \sum_{i<j} J_{(i,j)}^2})
   \biggr\rbrack   
\ee
where fluctuations of the $\sigma$s do influence the $J$s,
but 
\be \label{neur.ln_zre}
   \int_{-\infty}^\infty \prod_{i<j = 1}^N 
      \left( \sqrt{\frac{QN}{2\pi}} dJ_{(i,j)} \right)
    \ln \biggl\lbrack 
      \sum_{\{\sigma_i\}} 
      \exp{( \beta \sum_{i<j=1}^N J_{(i,j)}
                      \sigma_i \sigma_j
          )}
        \biggr\rbrack   
    \cdot  \exp{(-\frac{1}{2} Q N \sum_{i<j} J_{(i,j)}^2}) ,
\ee
or, in a shorthand notation,
\be 
   \int \cD J \;  \ln Z_\beta(J) \; \equiv \; 
   [[ \ln Z_\beta(J) ]] .
\ee
Usually one rewrites
\bea \label{neur.replica_trick}
   && \int \cD J \; \ln Z_\beta(J) \; = \; 
      \int \cD J \; \lim_{x\to 0} \frac{Z_\beta(J)^x-1}{x} 
   \nonumber \\
   && = \; \lim_{x\to 0} \int \cD J \; \frac{Z_\beta(J)^x-1}{x} 
      \; = \; \lim_{x\to 0} 
         \frac{\ln \lbrace 1+([[Z_\beta(J)^x]]-1)\rbrace}{x}
   \nonumber \\
   && = \; \lim_{x\to 0} \frac{\ln Z_{x\beta}^{\,\prime}}{x} .
\eea
For the second equality sign one has assumed that
$\int \cD J$ commutes with $\lim_{x\to 0}$,
in the third one that $\lim_{x\to 0} [[Z_\beta(J)^x]]=1$.
So far, $x$ as the ratio of two temperatures is real.
Rewriting the left hand side of (\ref{neur.replica_trick})
according to the right hand side is called the replica trick
\cite{zinn-justin}.
The uncontrolled approximation that usually enters the replica
trick is that now the right hand side is evaluated for
positive integer $x\equiv n$ and extrapolated to $n=0$.
Clearly a function that is known only for positive integer $n$
does not have a unique extrapolation to $n=0$ without further assumptions.
Nevertheless, this approximation is made, because it is
rather convenient. For integer $n$, 
$Z_\beta(J)^n$ is the partition function of an $n$ times replicated
system of which the logarithm is taken after the integration 
over the $J$s.
It is seen as follows.
We rewrite
\be\label{neur.zspin1}
  Z_\beta(J)^n \; = \; \sum_{\{\sigma_i^{(a)}\}}
  \exp{( \beta \sum_{a=1}^n \sum_{i<j=1}^N J_{(i,j)}
  \sigma_i^{(a)} \sigma_j^{(a)} )} ,
\ee
with $a=1,\dots , n$ labelling the replicated spin variables, so that
\bea\label{neur.zspin2}
  Z_{n\beta}^{\prime} & = &
  \int_{-\infty}^\infty \prod_{i<j = 1}^N dJ_{(i,j)} \cdot
  \sum_{\{\sigma_i^{(a)} = \pm 1\}} 
  \exp{(-S(J,\sigma^{(a)}))} , \nonumber \\
  S(J,\sigma^{(a)}) & = & 
  -\beta \sum_{a=1}^n \sum_{i<j=1}^N J_{(i,j)} \sigma_i^{(a)} \sigma_j^{(a)}
  + \frac{1}{2} Q N \sum_{i<j} J_{(i,j)}^2.
\eea
Linear terms in $\sigma$ and $J$ may be included according to
\be\label{neur.zspin3}
  S_{lin} \; = \; - h \sum_{a=1}^n \sum_{i=1}^N \sigma_i^{(a)}
  + c \sum_{i<j=1}^N J_{(i,j)} 
\ee
with constant external fields $h$ and $c$.

Apparently, because of integer $n$, 
$Z_{n\beta}^\prime$ has the form of models 
to which DLCE applies, with a hopping term 
\be\label{neur.hop}
  S_{hop}(J,\sigma^{(a)}) \; = \; - \beta \sum_{a=1}^n \sum_{i<j=1}^N
  J_{(i,j)} \sigma_i^{(a)} \sigma_j^{(a)},
\ee
a single link action
\be\label{neur.link}
 S^1(J_{(i,j)}) \; = \; c J_{(i,j)}
 + \frac{1}{2} Q N \; J_{(i,j)}^2,
\ee
and a single site action
\be\label{neur.site}
 S^\circ(\sigma_i^{(a)}) \; = \;
 - h \sum_{a=1}^n \sigma_i^{(a)}.
\ee
Depending on $n$ we distinguish the following cases.

\begin{itemize}
\item
$n=1$.
First we note that for $n=1$ we can directly apply DLCE to
$\ln Z_{\betap=\beta}^{\prime}$
and to derived quantities to obtain their series expansions
in $\beta$. 
But from a physical point of view, in a disordered system one
is not interested in $n=1$, because $n=1$ corresponds to the
completely annealed situation, in which the fast spins and the
slow interactions are in mutual equilibrium.
(In contrast, in particle physics one {\sl is} interested in the
$n=1$ case, cf.~our applications of DLCEs in the framework of
variational cumulant expansions of the $SU(2)$ Higgs model
\cite{DLCE}.)
\item
$n>1$, integer.
Again we apply DLCE to 
$\ln Z_{\betap=n\beta}^{\prime}$,
but have to account for the permutation symmetry between
the replicas. Formally, the replica symmetry plays a role
similar to an internal symmetry, e.g.~an $O(N)$ symmetry
in a scalar Higgs model. DLCEs with nontrivial internal symmetries
have been discussed in connection with the $SU(2)$ Higgs
model \cite{DLCE}.
Thus we can study "`unquenched"' equilibrium aspects of systems
with two temperatures and compare the
results from DLCEs adapted to ''internal'' replica symmetry
with Monte Carlo simulations for the same $n$
\cite{sherr1,janke}.
\item
$n=0$, the quenched limit.
As we will show in the next section, in order to discuss the
$x\to 0$ limit, we need not refer to $n$ times replicated
systems $Z_\beta(J)^n$ characterized by
(\ref{neur.hop})-(\ref{neur.site}),
but just to $Z_\beta(J)$ given by (\ref{neur.hop})-(\ref{neur.site})
with $n=1$. By means of special DLCEs , so-called quenched DLCEs
(QDLCEs), we directly calculate the left hand side of 
(\ref{neur.replica_trick}).
Therefore, setting $n=1$ in (\ref{neur.hop})-(\ref{neur.site})
in QDLCEs
does {\sl not} imply the completely annealed case, because we
first take the logarithm of $Z_\beta(J)$ and then average over the
$J$s.
\end{itemize}

%
%
\subsection{Avoiding the replica trick}

First
we adapt the notation to section 2 to include more general cases.
$\Lambda_0$ denotes the support of the spins, 
that is the set of lattice sites, with
$V=\vert\Lambda_0\vert$ denoting their total number.
$\overline\Lambda_1$ $\subseteq\Lambda_1$ are the pairs of sites
whose spins interact.
In accordance with (\ref{neur.zre}), we write for
the normalized link-average of a function $f(J)$
\be \label{sg.average}
   [[ f(J) ]] \; = \; \int \cD J \; f(J)
\ee
with
\bea
   \cD J & = & \prod_{l\in\overline\Lambda_1}
     d\mu(J(l)) \; ,
   \nonumber \\
   d\mu(J) & = & \cN_1 \; dJ \; \exp{(-S^1(J))} \; , \;
     \int_{-\infty}^\infty d\mu(J) = 1 .
\eea
It is convenient to introduce the single link expectation values
\be
  < g(J) >_1 \; \equiv \;
  \int d\mu (J) \; g(J)
\ee
and the generating function $W^1(I)$ by
\be
   \exp{W^1(I)} \; \equiv \;
   < \exp{(IJ)} >_1 .
\ee

The way in which the replica trick can be avoided
is examplified for the free energy density 
$W_{sp}/V$ of the spin system
averaged over the link couplings.
The partition function of the spin system
for a given distribution of the link interactions $J(x,y)$
is given by
\be
  \exp{W_{sp}(J)} \; = \;
  \cN_{sp} \; \int \cD\sigma \;
  \exp{(-S_{sp}(\sigma,J))},
\ee
where $W_{sp}(0)=0$ and
\bea
  S_{sp}(\sigma,J) & = &
  - \; \frac{1}{2} \sum_{x,y\in\Lambda_0} v(x,y)
   \sigma(x) \sigma(y) J(x,y), 
   \nonumber \\
  \cD\sigma & = & \prod_{x\in\Lambda_0}
   d\sigma(x) \cdot \exp{(-S^\circ(\sigma(x)))}.
\eea
Without loss of generality we
identify the support of the interaction
$v(x,y)=v(y,x)$ with the set 
$\overline{\Lambda}_1$ of lattice sites where 
$\cD J$ is supported,
\be
   \overline\Lambda_1 \; = \;
   \{ l=(x,y)\in\overline{\Lambda_0\times\Lambda_0} \; \vert \;
   v(x,y)\not= 0\}.
\ee
For simplicity we assume $v(x,y)$ to be of the form
(\ref{susc.vform}),
so that $K$ is a measure of the strength of the interactions
$v(x,y)$.

\vskip7pt

The free energy density of the spin system allows for a
series expansion in the standard LCE sense, 
with the link field $J(l)$ playing the role
of a "background field",
\be \label{sg.wsp_series}
   \frac{1}{V} W_{sp}(J) \; = \;
   \sum_{L\geq0} (2K)^L 
   \sum_{\Gamma\in\cG^{sp}_0(L)}w^{sp}(\Gamma,J).
\ee
Here $\cG_E^{sp}(L)$ (with $E=0$) denotes the
set of equivalence classes of connected LCE graphs
with $E$ external lines and $L$ internal lines.
The spin-weights $w^{sp}(\Gamma,J)$ are of the form
\be \label{sg.wsp_weight}
 w^{sp}(\Gamma,J) \; = \; R^{sp}(\Gamma)
  \sum_{\cL_\Gamma\to\overline{\Lambda}_1}^{\quad\prime}
  \prod_{l\in\overline{\Lambda}_1}
  J(l)^{m(l)} .
\ee
The sum is taken over all non-vanishing lattice embeddings
of the graph $\Gamma$.
It runs over all maps of internal lines of the graph
$\Gamma$ to pairs of lattice sites of $\overline\Lambda_1$
that are consistent with the graph topology in the
sense discussed in section 3.
For every $l\in\overline\Lambda_1$, $m(l)$ denotes the number
of lines of $\Gamma$ that are mapped onto the link $l$
by the embedding.
All other factors that contribute to the weight
are collected in the prefactor $R^{sp}(\Gamma)$,
including the inverse topological symmetry number of $\Gamma$.

\vskip7pt

Next we want to express
$[[W_{sp}(J)]]$ as a series in $K$ by means of DLCE.
Toward this end we set $f(J)=W_{sp}(J)$ and insert the series
(\ref{sg.wsp_series}) with (\ref{sg.wsp_weight})
into (\ref{sg.average}).
At this stage we are not concerned with question of 
(uniform or dominated)
convergence and obtain
\bea \label{sg.wspint}
  [[ \frac{1}{V} \; W_{sp}(J) ]] & = &
   \sum_{L\geq 0} (2K)^L \sum_{\Gamma\in\cG_0^{sp}(L)}
    \int\cD J \; w^{sp}(\Gamma,J)
   \nonumber  \\
  & = & \sum_{L\geq 0} (2K)^L \sum_{\Gamma\in\cG_0^{sp}(L)}
   R^{sp}(\Gamma)  \sum_{\cL_\Gamma\to\overline{\Lambda}_1}^{\quad\prime}
  \prod_{l\in\overline{\Lambda}_1}
  <J(l)^{m(l)}>_1 .
\eea

The next step is to express the full single link expectation values
in terms of the connected ones. They are related by
\be \label{sg.conn}
   < J^m >_1 \; = \;
   \sum_{\Pi\in\cP(\underline{m})} \prod_{P\in\Pi}
   < J^{|P|} >_1^c,
\ee
where $\cP(\underline{m})$ denotes the set of all partitions
of $\underline{m}=\{1,\dots ,m\}$ into non-empty,
mutually disjoint subsets of
$\underline{m}$. 
$|P|$ is the number of elements of the set $P$.
The relation (\ref{sg.conn})
is equivalent to the partition of all lines of $\Gamma$ that are
mapped to the same lattice link into multiple lines,
with every multiple line contributing a factor
\be \label{sg.wdiff}
   < J^{|P|} >_1^c \; = \;
   \left. \frac{\partial^{|P|} W^1(I)}{\partial I^{|P|}}
   \right\vert_{I=0} \; = \;
   m^{1 c}_{|P|}.
\ee
Using (\ref{sg.conn}), (\ref{sg.wdiff}) we rewrite
(\ref{sg.wspint}) as
\be \label{sg.wsptodlce}
  [[ \frac{1}{V} \; W_{sp}(J) ]]  =
   \sum_{L\geq 0} (2K)^L \sum_{\Gamma\in\cG_0^{sp}(L)}
    R^{sp}(\Gamma) \;
    \sum_{\Pi\in\cP(\cL_\Gamma)}
   \; \left( \prod_{P\in\Pi} m^{1 c}_{|P|} \right)
   \;
  \left( \sum_{\Pi\to\overline{\Lambda}_1}^{\quad\prime}
  \prod_{l\in\overline{\Lambda}_1}
   1 \right) .
\ee
The last summation in (\ref{sg.wsptodlce}) is over all
maps $\cL_\Gamma\to\overline\Lambda_1$ of the lines
of $\Gamma$ to the lattice links of $\overline\Lambda_1$
subject to the constraint that all lines that belong
to the same multiple-line corresponding to some $P\in\Pi$
are mapped onto the same lattice link.

Finally we rewrite (\ref{sg.wsptodlce}) as a sum over
\mlgraphs.
To this end, we first observe that for every
$\Gamma\in\cG^{sp}_0(L)$, every partition
$\Pi\in\cP(\cL_\Gamma)$ of the lines of $\Gamma$ into 
multiple-lines generates a \mlgraph $\Delta=(\Gamma,\Pi)$ in the
obvious way.
Let us denote by
$\overline\cG_{0,0}(L)$ the subset of \mlgraphs of
$\cG_{0,0}(L)$ that stay connected after decomposition
of all multiple lines.
(These are the \mlgraphs which stay connected in the usual
graph theoretical sense, when the dashed lines are omitted.)
For every $\Delta\in\overline\cG_{0,0}(L)$ there is a unique
$\Gamma(\Delta)\in\cG_0^{sp}(L)$ and at least one
$\Pi\in\cP(\cL_{\Gamma(\Delta)})$ such that
$(\Gamma(\Delta),\Pi)=\Delta$.
Let $n_\Delta$ be the (uniquely determined) number of
partitions $\Pi\in\cP(\cL_{\Gamma(\Delta)})$ with
$(\Gamma(\Delta),\Pi)=\Delta$,
and $\Pi(\Delta)$ such an arbitrary partition.
Eq.~(\ref{sg.wsptodlce}) then becomes
\be \label{sg.wdlce}
  [[ \frac{1}{V} \; W_{sp}(J) ]]  =
   \sum_{L\geq 0} (2K)^L \sum_{\Delta\in\overline\cG_{0,0}(L)}
    n_\Delta R^{sp}(\Gamma(\Delta)) \;
   \; \left( \prod_{P\in\Pi(\Delta)} m^{1 c}_{|P|} \right) \;
  \left( \sum_{\Pi(\Delta)\to\overline{\Lambda}_1}^{\quad\prime}
   1 \right) .
\ee
The last bracket of (\ref{sg.wdlce})
is the lattice embedding factor of the
\mlgraph $\Delta$.
The second bracket from the right does not depend on the choice
of $\Pi(\Delta)$ and is the product of the multiple-line
coupling constants as defined in section 3.
Finally, $n_\Delta R^{sp}(\Gamma(\Delta))$ is precisely
the remaining part of the weight of $\Delta$
that was described in detail in section 3, endowed
with the correct inverse topological symmetry number
of the \mlgraph $\Delta$ (because of the factor $n_\Delta$).

In summary, we obtain the series expansion of the link-averaged
free energy density in terms of DLCE graphs,
\be \label{sg.wseries}
    [[ \frac{1}{V} \; W_{sp}(J) ]] \; = \;
    \sum_{L\geq 0} (2K)^L 
     \sum_{\Delta\in\overline{\cG}_{0,0}(L)} w(\Delta) .
\ee
The weight $w(\Delta)$ of a \mlgraph $\Delta$ is defined
and computed according to the rules given in section 3.

Eq.~(\ref{sg.wseries}) is the series representation of the
link-averaged free energy density of the spin system,
i.e.~the free energy density of the $n=0$ replica system,
in terms of DLCE graphs.
It looks much like the series representation of the
$1$-replica system, which is given by
\be \label{sg.w1series}
    \frac{1}{V} \; W_{1-repl} \; \equiv \;
    \frac{1}{V} \; \ln{[[ \exp{W_{sp}(J)} ]]} \; = \;
    \sum_{L\geq 0} (2K)^L 
     \sum_{\Delta\in\cG_{0,0}(L)} w(\Delta)
\ee
according to the discussion of section 2.
We recall that
$\cG_{0,0}(L)$ is the set of DLCE vacuum graphs with $L$ bare lines
that are connected in the generalized DLCE sense.
Comparing (\ref{sg.wseries}) and (\ref{sg.w1series}),
the transition from $n=1$ to $n=0$ replicas is achieved by
keeping only the subset
$\overline{\cG}_{0,0}(L)\subseteq\cG_{0,0}(L)$ of
\mlgraphs that are connected in the original (LCE) sense.

We emphasize that the restriction of DLCEs to QDLCEs is not an ad hoc
(or intuitively motivated) assumption but a derived consequence
of the fact that the logarithm is taken before the integration
$\int \cD J$.
This procedure accounts for {\sl all} graphs that contribute to a
given order in $K$. Thus we do have to truncate the series unless
the series can be completely summed up, as it happens in
exceptional cases. 

We expect that the series (\ref{sg.wseries}) are convergent
for a large class of interactions $S^1(J)$ and $v(x,y)$
if the coupling constant $K$ is sufficiently small.
For special interactions most of the \mlgraphs yield 
vanishing contributions so that we can further restrict
the sum to a subset of $\overline{\cG}_{0,0}(L)$.
An example is given by the mean field type of interaction
of the Sherrington-Kirkpatrick model,
cf.~Sect.~5.2.

%
%
\subsection{Applications of QDLCEs}

In the following we list some examples for systems of which we can
study the phase structure by means of QDLCEs.
Their actions are special cases of 
(\ref{neur.hop})-(\ref{neur.site})
with $n=1$ (as explained above) and the following choice of variables.
\begin{itemize}
\item
Infinite range models.
Choose $J_{(i,j)}\in\mathbb{R}$ as before, $\sigma_i\in{\pm 1}$,
$i,j\in\{1,\dots ,N\}$,
\bea
v(x,y) & = & K (1-\delta_{x,y}),  \nonumber \\
\overline\Lambda_1 & \equiv & \Lambda_1 \;
\mbox{is the set of all pairs
of sites,} \\
S^1(J) & = & N \; \frac{1}{2} \; J^2 . \nonumber
\eea
Now the sum over the sites in (\ref{neur.hop}) 
runs over arbitrary pairs $(i,j)$, $i<j$, and
we obtain the infinite range Sherrington-Kirkpatrick model.
For infinte range and in the thermodynamic limit
($N\to\infty$), the phase structure can be solved by replica
mean field theory, cf.~e.g.~\cite{fischer}.
For QDLCEs the infinite range and $N\to\infty$ limits imply that
only tree graphs of 2-lines contribute to the series
of the free energy density, such as
\bea 
   && \nonumber \\
   && \nonumber \\
   \\
   && \begin{picture}(15.0,15.0)
         \epsfig{bbllx=237,bblly=333,
                 bburx=717,bbury=253,
                 file=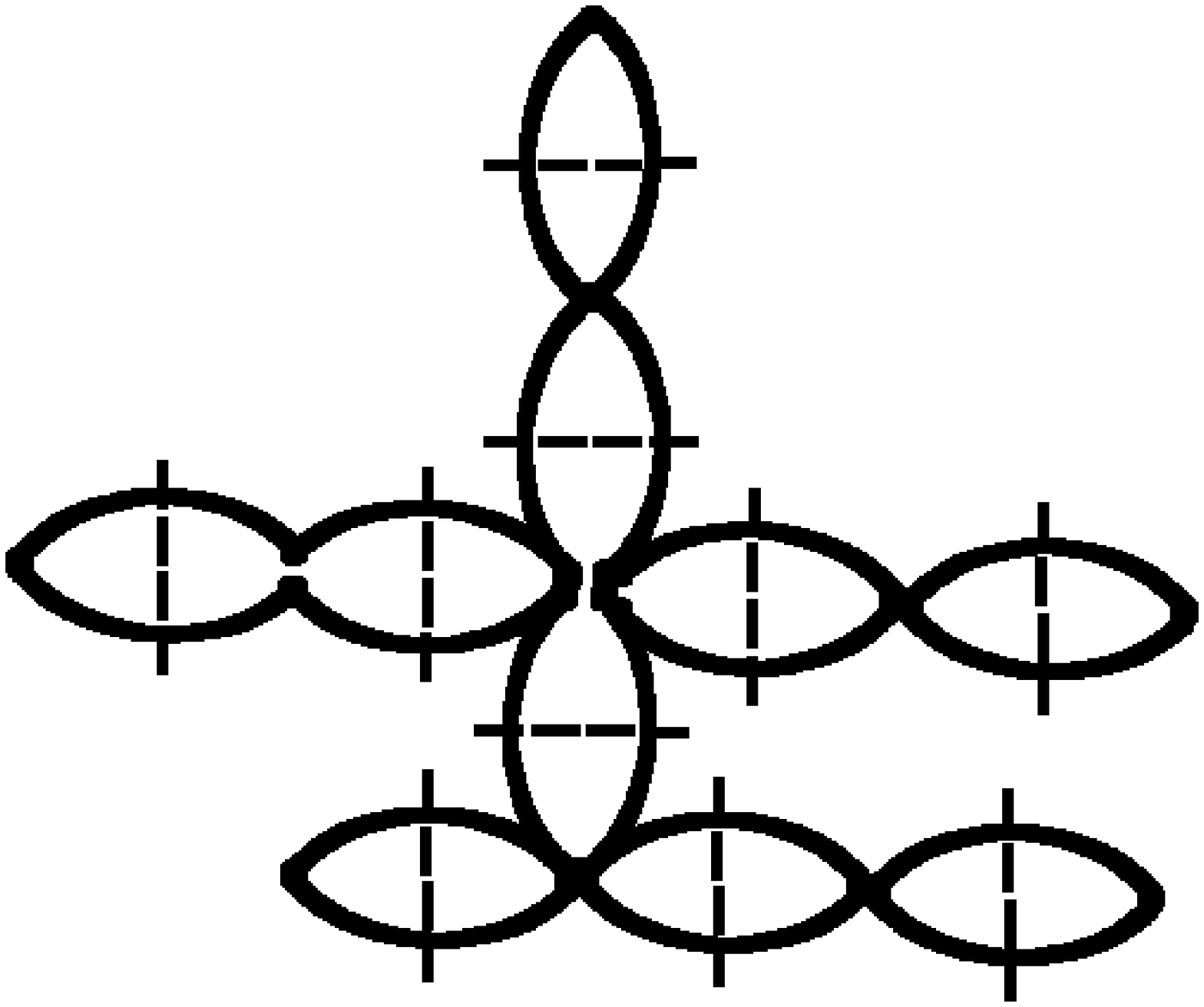,
                 angle=2,
                 scale=0.22}
     \end{picture} 
   \nonumber \\ 
   && \nonumber \\
   && \nonumber
\eea
The reason for that is that each 2-line gets a factor of $1/N$ from the
$S_1$-part of the action, but each vertex gets a factor $(N-1)$ 
from the embedding onto a lattice
$\Lambda_0$. 
The contribution of every tree graph to the free energy is
proportional to $N$.
If a chain of 2-lines connecting the vertices
gets closed, forming a loop, there is one $(N-1)$ less
in the total embedding factor. Thus the contribution is suppressed
by $1/(N-1)$ for every loop and vanishes in the thermodynamic limit.
Because of the simple tree structure there is a chance for summing
up the series. This is currently under investigation.

\item
Finite range connectivity.
The sum $\sum_{i<j}$ of the spins is now restricted to next-neighbours
or, more generally, to a finite number of pairs.
Rather than specifying $S_1(J_{(i,j)})$ of (\ref{neur.link}),
it is sufficient for DLCEs and QDLCEs to choose
$\exp (-S_1(J_{(i,j)}))$.
Let
\be
   exp{(-S_1(J))} \; = \; (1-p) \delta(J) + p \delta(J-1)
\ee
with $p\in [0,1]$. The variables
$J_{(i,j)}\in\{0,1\}$ can then be interpreted 
as occupation numbers of the bonds.
Furthermore, if we choose
$\sigma_i\in\{\pm 1\}$ we obtain a

\begin{itemize}
\item
bond-diluted Ising model. 

Choosing 
$\sigma_i\in\mathbb{Z}_q$ we obtain a 
\item
bond-diluted $q$-state Potts model.

If $\sigma_i\in S_q$, we obtain a 
\item
bond-diluted Heisenberg model.
\end{itemize}

\end{itemize}

QDLCEs provide a systematic analytic expansion for disordered
systems with bond dilution in a quenched limit.
Coming from the high temperature (small $\beta$) region one can
study the phase structure as a function of the degree of dilution.
Work in this direction is in progress.

%
%
%
%
\section{Summary and Conclusions}

In this paper we have introduced a new expansion scheme for 3-point
interactions or, more precisely, for point-link-point interactions.
This scheme generalizes linked cluster expansions
for 2-point interactions 
by including hopping parameter terms endowed with their own dynamics.
In chapters 3-4 we have developed a \mlgraph theory with an
additional new type of multiple-line connectivity.
We have introduced appropriate equivalence classes of graphs
and discussed the issue of renormalization.
These notions are required for
an algorithmic generation of graphs.
Because of
the fast proliferation of graphs already at low orders in the expansion,
a computer aided implementation becomes unavoidable,
if one is interested in higher orders of the expansion than 
we have computed so far.

In Sect.~5 we have shown how to avoid the replica trick for
calculating the free energy of disordered systems in the
quenched limit.
DLCEs are a systematic expansion method to study the phase structure 
of disordered systems.
It is systematic in the sense that we do not restrict the expansion to
certain subclasses of graphs that can be summed up,
but we identify and keep {\sl all} graphs that contribute to a
given order in the expansion parameter.
DLCEs provide an analytic tool for studying systems in situations
in which it has been impossible so far.

\section*{Acknowledgment}

We would like to thank Reimar K\"uhn (Heidelberg) for discussions.

%
%
%
%



\begin{thebibliography}{9}

\bibitem{wortis} M.~Wortis, "Linked cluster expansion",
in Phase transition and critical phenomena, vol.3, eds.
C.~Domb and M.S.~Green (Academic Press,London 1974).
%
\bibitem{itzykson} C.~Itzykson, J.-M.~Drouffe,
"Statistical field theory", vol.2, Cambridge University Press, 1989.
%
\bibitem{guttmann} A.J.~Guttmann,
"Asymptotic analysis of Power-Series Expansions",
in Phase transition and critical phenomena, vol.13, eds.
C.~Domb and J.L.~Lebowitz (Academic Press).
%
\bibitem{LW1} M.~L\"uscher and  P.~Weisz,
Nucl.~Phys. {\bf B300}[FS22] (1988) 325.
%
\bibitem{thomas1} T.~Reisz,
Nucl.~Phys. {\bf B450} (1995) 569.
%
\bibitem{thomas2} T.~Reisz,
Phys.~Lett. {\bf 360B} (1995) 77.
%
\bibitem{campostrini} M.~Campostrini, A.~Pelissetto, P.~Rossi
and E.~Vicari,
Nucl.~Phys. {\bf B459} (1996) 207.
%
\bibitem{zinn} S.~Zinn, S.-N.~Lai and M.~E.~Fisher,
Phys.~Rev. {\bf E54} (1996) 1176.
%
\bibitem{butera} P.~Butera and N.~Comi,
Phys.~Rev. {\bf E55} (1997) 6391.
%
\bibitem{hilde1} H.~Meyer-Ortmanns and T.~Reisz,
J.~Stat.~Phys. {\bf 87} (1997) 755.
%
\bibitem{sherrington} D.~Sherrington and S.~Kirkpatrick,
Phys.~Rev.~Lett {\bf 35} (1975) 1972.
%
\bibitem{dotsenko} V.~Dotsenko, S.~Franz and M.~M\'ezard,
J.~Phys. {\bf A27} (1994) 2351.
%
\bibitem{penney}
R.~W.~Penney, A.~C.~C.~Coolen, and D.~Sherrington, 
J.~Phys.~A: Math.~Gen. {\bf 26} (1993) 3681.
A.~C.~C.~Coolen, R.~W.~Penney, and D.~Sherrington, 
Phys.~Rev. {\bf B48} (1993) 16116.
%
\bibitem{DLCE}  H.~Meyer-Ortmanns and T.~Reisz,
Int.~J.~Mod.~Phys. {A14} (1999) 947.
%
\bibitem{thomas3} T.~Reisz,
Nucl.~Phys. {\bf B527} (1998) 363.
%
\bibitem{fischer}
K.~H.~Fischer and J.~A.~Hertz,
``Spin Glasses'', Cambridge University Press, 
Cambridge 1991.
%
\bibitem{zinn-justin} J.~Zinn-Justin, 
``Quantum Field Theory and Critical Phenomena'', 3rd Edition,
Oxford Science Publication, Clarendon Press, Oxford 1996.
%
\bibitem{sherr1} D.~Sherrington, 
J.~Phys. {\bf A13} (1980) 637;
I.~Kondor, J.~Phys. {\bf A16} (1983) L127.
%
\bibitem{janke} B.~Berg and W.~Janke, 
Phys.~Rev.~Lett {\bf 80} (1998) 4771
%
\end{thebibliography}
\end{document}